\documentclass[10pt,letterpaper]{article}
\usepackage[top=0.85in, lmargin = 1in] {geometry} %% Added by Hannah

\newcommand{\beginsupplement}{%
        \setcounter{table}{0}
        \renewcommand{\thetable}{S\arabic{table}}%
        \setcounter{figure}{0}
        \renewcommand{\thefigure}{S\arabic{figure}}%
        \setcounter{subsection}{0}
        \renewcommand{\thesubsection}{S\arabic {subsection} Text}%
     } %% Added by Hannah

\usepackage{color}%% Added by Hannah

\usepackage{changepage}

\usepackage[utf8]{inputenc}

\usepackage{textcomp,marvosym}

\usepackage{fixltx2e}

\usepackage{amsmath,amssymb}

\usepackage{cite}

\usepackage{nameref,hyperref}

\usepackage[right]{lineno}

\usepackage{microtype}
\DisableLigatures[f]{encoding = *, family = * }

\usepackage{rotating}

\raggedright
\usepackage[aboveskip=1pt,labelfont=bf,labelsep=period,justification=raggedright,singlelinecheck=off]{caption}

\makeatletter
\renewcommand{\@biblabel}[1]{\quad#1.}
\makeatother

\date{}

\usepackage{lastpage,fancyhdr,graphicx}
\usepackage{epstopdf}
\fancyheadoffset[L]{2.25in}
\fancyfootoffset[L]{2.25in}
\newcommand{\Ecomment}[1]{\textcolor{cyan}{}}
\newcommand{\Hcomment}[1]{\textcolor{blue}{}}
\newcommand{\Acomment}[1]{\textcolor{green}{}}

\newcommand{\Ecommentnew}[1]{\textcolor{green}{}}

\usepackage{ulem}
\usepackage{soul,xcolor}
\newcommand\Hcheck{\bgroup\markoverwith{\textcolor{blue}{\rule[0.5ex]{2pt}{0.4pt}}}\ULon}
\begin{document}
\vspace*{0.35in}

% Please capitalize all terms in the title except conjunctions, prepositions, and articles.
\begin{flushleft}
{\Large
\textbf\newline{Predictive coding in area V4: dynamic shape discrimination under partial occlusion}
}
\newline
\\
Hannah Choi\textsuperscript{1,2,3*},
Anitha Pasupathy\textsuperscript{2,3,4},
Eric Shea-Brown\textsuperscript{1,3}
\\
\bigskip
\bf{1} Department of Applied Mathematics, University of Washington, Seattle, WA, USA
\\
\bf{2} Department of Biological Structure, University of Washington, Seattle, WA, USA
\\
\bf{3} UW Institute for Neuroengineering, University of Washington, Seattle, WA, USA
\\
\bf{4} Washington National Primate Research Center, University of Washington, Seattle, WA, USA
\bigskip

% Primary Equal Contribution Note
% \Yinyang These authors contributed equally to this work.

% Additional Equal Contribution Note
% Also use this double-dagger symbol for special authorship notes, such as senior authorship.
% \ddag These authors also contributed equally to this work.

% \textcurrency a Insert current address of first author with an address update
% \textcurrency b Insert current address of second author with an address update
% \textcurrency c Insert current address of third author with an address update

% Group/Consortium Author Note
% \textpilcrow Membership list can be found in the Acknowledgments section.

* hannahch@uw.edu

\end{flushleft}
\section*{Abstract}
% Try to follow the following format for Abstract -- question/goal, methods, results and conclusion.

The primate visual system has an exquisite ability to discriminate partially occluded shapes. Recent electrophysiological recordings suggest that response dynamics in intermediate visual cortical area V4, shaped by feedback from prefrontal cortex (PFC), may play a key role.  To probe the algorithms that may underlie these findings, we build and test a model of V4 and PFC interactions based on a hierarchical predictive coding framework. We propose that probabilistic inference occurs in two steps.  Initially, V4 responses are driven solely by  bottom-up sensory input and are thus strongly influenced by the level of occlusion.  After a delay, V4 responses combine both feedforward input and feedback signals from the PFC; the latter reflect predictions made by PFC about the visual stimulus underlying V4 activity.\Hcomment{Revised, I think if we just say ``underlying visual stimulus'' instead of ``visual stimulus underlying V4 activity'' it may sound like PFC makes direct predictions on the visual stimulus disregarding the whole hierarchical computation.}\Hcomment{PFC predictions are indeed compared to V4 activity- In predictive coding, higher areas make predictions on lower area activity.}\Acomment{Is the prediction about V4 activity or the identity of the underlying visual stimulus?}\Acomment{I disagree. Its a prediction of what V4 activity represents. Its a prediction of what the stimulus was that gave rise to that V4 activity. You don't need to predict V4 activity. You see the V4 activity.} We find that this model captures key features of V4 and PFC dynamics observed in experiments.  Specifically, PFC responses are strongest for occluded stimuli and delayed responses in V4 are less sensitive to occlusion, supporting our hypothesis that the feedback signals from PFC underlie robust discrimination of occluded shapes. Thus, our study proposes that area V4 and PFC participate in hierarchical inference, with  feedback signals encoding top-down predictions about occluded shapes.
\Acomment{i've rewritten the abstract a bit. See if you like it}

% Please keep the Author Summary between 150 and 200 words
%Aim to highlight where your work fits within a broader context; present the significance or possible implications of your work simply and objectively; and avoid the use of acronyms and complex terminology wherever possible. The goal is to make your findings accessible to a wide audience that includes both scientists and non-scientists.
\section*{Author Summary}
We can easily recognize objects even when they are partially occluded, but how our brains achieve this is an open question.  While many previous models focus on feedforward computations, there is an abundance of feedback connectivity in the ventral visual pathway whose functional role in processing occluded shapes is relatively unexplored. Here we undertake a computational study that contributes to closing this gap.  Given recent experimental evidence for shape-discriminating signals under occlusion in the intermediate visual area V4 and prefrontal cortex, we focus on the interplay of feedforward and feedback signals in V4. We introduce a novel interpretation of feedback in shape recognition in terms of network model that implements predictive coding, where the feedback represents the predictions made by prefrontal cortex about V4 activity. \Acomment{unclear what this previous sentence means. Simplify. What is novel? the network model or that the model implements predictive coding or that feedback represents predictions?\Hcomment{Both, but implementation of predictive coding is novel} \Acomment{The sentence is unclear to me. Its a set of words that I find hard to parse. Please simplify and clarify} DOn't understand. Also, predictions of PFC are about the underlying stimulus and not about V4 activity, correct? \Hcomment{PFC predictions are compared to V4 activity, as the system tries to minimize the difference between V4 activity and PFC predictions. Because the connection weights are trained on unoccluded shapes, the perfect prediction on V4 activity is made when the input shapes are unoccluded. In that sense, PFC makes prediction also on the underlying stimulus.}}We propose that V4 integrates both the feedforward sensory signals and feedback predictions to obtain an optimal representation of neuronal responses. Our results suggest that the predictive feedback signals increase shape discriminability under partial occlusion, addressing a possible algorithmic role of feedback in area V4.

%\linenumbers

\section*{Introduction}

In natural scenes, objects rarely appear in isolation; rather, animals often have to discriminate and recognize partially occluded objects. While recognition under occlusion is difficult for even the best computer vision system, animals seldom have trouble. But the neural basis of this capacity is poorly understood. 

Feedback projections from higher cortices are hypothesized to be important for successful recognition of occluded objects \cite{Rust10, Gregoriou14}, and there are abundant feedback connections in the visual stream.  Despite this, models of object recognition are typically  hierarchical feedforward circuits \cite{Fukushima80, Riesenhuber99, Serre07, Cadieu07, Yamins14}.  This is partly because of the complexity of including feedback signals, but also because little is known about where the relevant feedback signals originate, where they terminate in visual cortex, and how they contribute to recognition. Developing a computational framework explaining how feedback facilitates shape recognition under occlusion, therefore, is a prominent challenge for visual neuroscience. \Hcomment{Revised.}\Acomment{don't know what the point of this last sentence is. It seems rather general and i'm not sure. I might just say instead that in this study you develop a computational framework to understand how feedback facilitates recognitino under occlusion.}

Recent experimental results provide key insights into how interactions between area V4, a fundamental stage in the primate shape processing pathway \cite{Roe12,Pasupathy99,Pasupathy01}, \Hcomment{Done}\Acomment{I might cite something more appropriate like ROe, Chelazzi, Connor, et al, 2012 Neuron} and the prefrontal cortex, important for the control of complex behavior \cite{Miller01}\Hcomment{Done}\Acomment{cite Miller and Cohen, 2001}, may underlie the ability to recognize partially occluded objects \cite{Kosai14, Pasupathy15}. Specifically, in monkeys trained to discriminate pairs of shapes under varying degrees of occlusion, dynamics of V4 and PFC activity suggest  that feedback signals from PFC to area V4 may serve to discount the effect of occlusion on the responses of V4 neurons -- thereby increasing shape selectivity.   This raises the question of how the feedback signals in V4-PFC circuitry perform the computation necessary for shape recognition. In this paper, we propose and test the hypothesis that this occurs via a hierarchical predictive coding. 

Predictive coding has been proposed as a method to create efficient neural codes, and has successfully described neural responses in a variety of different sensory systems \cite{Bogacz15, Bastos12, Friston09a, Friston09b, Srinivasan82, Rao99a, Spratling16, Rao97,Rao99b, Rao04, Rao05, Lee03, Yuille06}.  Notably, the predictive coding framework reproduces center-surround antagonism in retina \cite{Srinivasan82} and endstopping effects in V1 \cite{Rao99a}. In these studies, feedforward signals from each cortical area  represent the residual errors between the feedback predictions and the encoding expectation. \Hcomment{Actual cortical activity is the maintained neuronal state- Friston (Bastos et al 2012) calls it either encoding expectation or conditional expectation. I changed it to the term ``encoding expectation'' here- maybe that's less confusing}\Acomment{I don't understand this preceding sentence. What is actual cortical acitivty? You don't mean the feedforward signal from any area, I think you mean the output from any area? Otherwise your terminology here and in the last para are inconsistent} This interpretation of feedforward signals, however, has met the criticism\cite{Koch99} that it implies reduced firing  when familiar sensory inputs are encountered, differing from the common view in which sensory neurons respond strongly to preferred features.  Here, we introduce a novel implementation of predictive coding, where the responses in V4 and PFC correspond to their most likely (or optimal) values given the stimulus and a hierarchical representation of its likelihood. Furthermore, the hierarchical inference is implemented in two steps, initially reflecting only the feedforward sensory signals and later integrating the feedback predictions, to explain the dynamic shape selective responses in V4. 

In addition to assigning an algorithmic role to the feedback signals, our model makes further predictions on the structure of the network, representation of the stimuli, and prior expectations encoded in V4 and PFC.  Previous studies have shown that shapes can be discriminated based on V4 activity at the population level (\cite{Meyers08,Pasupathy02}), and shape identity information is already available at the level of V4. However, in our model feedback predictions effectively re-map the population responses \Hcomment{The information of occlusion level encoded by V4 unit 3 is available to V4 unit 1 and 2, thus the information is ``re-mapped''.}\Acomment{why remap here?}and amplify the shape identity information that are reduced by partial occlusion. %Moreover, the model does not require additional mechanisms for segmentation of the occluders from the shapes as in previous models of pattern and shape recognition under occlusion \cite{Fukushima87, Fukushima01, Fukushima05}
%Our model, therefore, not only captures characteristics of neural responses characteristics in experiments, but also extends hierarchical Bayesian inference approach to decision making on shape identity, based on probabilistic population codes \cite{Beck08, Drugowitsch12}. 

In sum, our model suggests that feedback signals to V4 during the representation of occluded shapes can be interpreted in the context of predictive coding. These results shed light on how prior expectation contribute to the recognition of complex images in V4 and higher cortical areas. \Acomment{Rewrote this. See if you like it}

\section*{Materials and Methods}

% For figure citations, please use "Fig." instead of "Figure".

\subsection*{Experiments}

Experimental procedures are described in detail by \cite{Kosai14,Pasupathy15}, and are only briefly outlined in this section to provide the background. 

Animals were trained on a sequential shape discrimination task, where two stimuli were presented in sequence and the animal had to report whether they were the same or different with a rightward or a leftward saccade, respectively. The second stimulus in the sequence was presented in the receptive field of the V4 neuron under study and was partially occluded. During recordings in area V4, all task details were customized to the preferences of the single neuron under study. Specifically, one of the two discriminanda was a preferred shape that elicited strong responses from the neuron while the other was a non-preferred shape. Both shapes were presented in a preferred color for the cell and the occluding dots were in a non-preferred color so they provided only a modulatory influence.  For recordings in the PFC, we studied many neurons simultaneously and did not customize stimulus shape or color to individual neuronal preferences as is customary in the field. Each day the experimental session began as follows. We chose two stimuli to serve as the discriminanda. This was followed by two phases. First, during the training phase, animals performed the sequential discrimination task with the unoccluded versions of the discriminanda. This typically included 20 attempts and was to ensure that the the unoccluded versions of the discriminanda were discriminable in the periphery. This was followed by the test phase during which the discriminanda were occluded to different levels. 

\subsection*{Coding assumptions}

We explain the response dynamics of V4 and PFC neurons during the shape discrimination task by building a computational model based on a few coding principles, which we introduce here. 

First, we assume that average firing rates of the neuronal populations recorded in experiments reflect the most likely representation of the neuronal responses given the input visual stimulus and a specific hierarchical model of the responses that we define below. Thus, assuming the sensory system seeks to infer the most likely representation of neuronal responses  $\{r_1,...,r_n\}$ of hierarchical areas ranging from the lowest area 1 to the highest area n, we simply find the set of responses that maximizes the posterior probability $p(r_1,...,r_n|\kappa)$, where $\kappa$ represents the sensory input.  We refer to these as the optimal firing rates.   \Ecomment{Tried to streamline / compress this.}

\Hcomment{I think I will just move and merge this paragraph (commented out) with the next section (Model architecture)}
Second, the model is constructed based on the hierarchical predictive coding principle. In predictive coding \cite{Rao99a,Friston09a,Bogacz15}, feedback from higher cortical areas is interpreted as a prediction about activities in lower cortical areas. In the lower cortical areas,  bottom-up sensory signals  are combined with these top-down predictions.  With the predictions and the sensory inputs thus combined, probability distributions of the neural responses are constructed based on hierarchical Bayesian inference \cite{Rao99a,Bogacz15,Lee03,Yuille06}. Under this assumption, combined with predictive coding, neuronal activities depend on the activities of the next higher area, but are independent of activities in other cortical areas. In other words, the neurons in area $i+1$, whose activity is denoted as $r_{i+1}$, make the `top-down' prediction $Pred(r_{i+1})$ of the neuronal activity $r_{i}$ in area $i$. The noise $\eta_i$ characterizing the differences between the actual neuronal response $r_{i}$ and the prediction made by the next higher layer $Pred(r_{i+1})$, is  given as
\begin{align}
\eta_i = r_{i}-Pred(r_{i+1}).
\end{align}
We assumed the noises to have a distribution $g_i(\eta_i)$ with zero mean.  This leads to  $p(r_{i}|r_{i+1})$, 
the distribution of the neuronal activity $r_{i}$ in area $i$ given the next level activity  $r_{i+1}$, having its mean at the top-down prediction $Pred(r_{i+1})$.  \Ecommentnew{rewrote this. please make sure right or fix as needed}

The posterior probability of the response representation across all levels given the sensory stimulus $\kappa$ therefore factors as 
\begin{align}
\begin{split}
p(r_1,...,r_n|\kappa) &= \nu\cdot p(\kappa|r_1,...,r_n)p(r_1,...,r_n)\\
&= \nu\cdot p(\kappa|r_1)p(r_1|r_2)...p(r_{n-1}|r_n)p(r_n),
\end{split}
\end{align}
\noindent where $\nu$ is a normalization constant.

Above we described the general and classical framework for hierarchical representation of a stimulus $\kappa$ via a sequence of firing rates. In summary, we assume that the brain aims to have neuronal activity in every layer get as close as possible to the prediction made by the responses of the next higher layer, where the discrepancy is given by a noise term $\eta_i$. Then, the neural firing rates adjust to those that are most consistent, i.e., most likely, given the stimulus $\kappa$. We next describe the specific form of the representation that we use here. 

\subsection*{Model architecture}

\begin{figure}[!htb]%[h]
\noindent \centering
\includegraphics[scale=0.65]{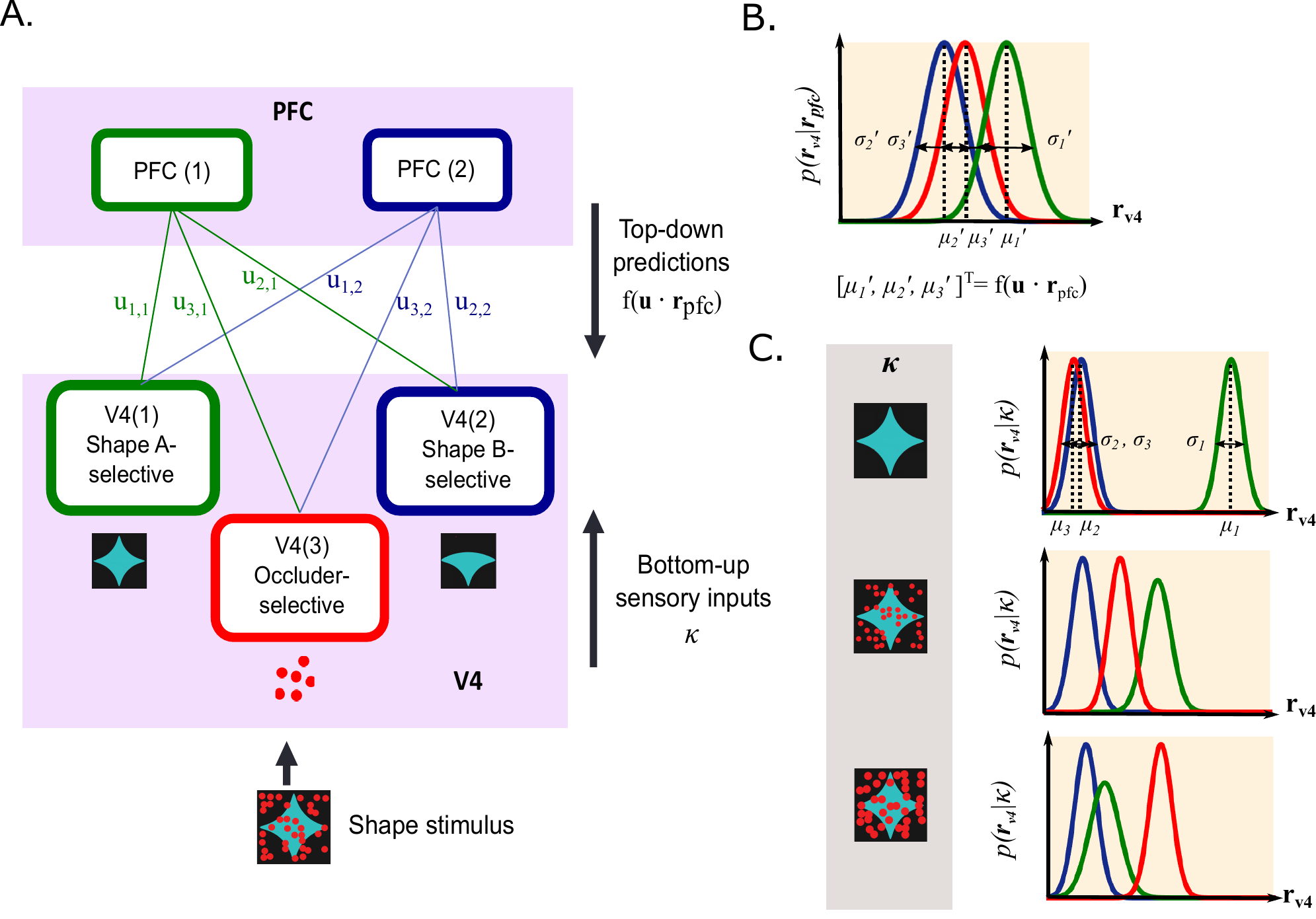}
\caption{{\bf Schematic diagram of network model.}
(A) Model network of V4 and PFC populations and the schematic of the input shape stimulus. Note that the model is not image computable, and the input stimulus in the figure is given to illustrate the model setup.  (B) Top-down predictions made by PFC on each of the three V4 units are represented by Gaussian distributions with  means at $f(\bf{u}\cdot\bf{r_{pfc}})=\bf{u}\cdot\bf{r_{pfc}}$. (C) Bottom-up component, which is represented by the conditional probability distributions of the V4 responses given the shape stimulus. When the input stimulus is unoccluded shape A, the response distribution of the shape A-selective V4 population has a higher mean than those of the shape B- and occluder-selective populations. As the occlusion level increases, the mean of the shape A-selective response distribution decreases and the standard deviation increases.  Shape B-selective distribution stays at the constant baseline and the occluder-selective response distribution moves towards higher rates. The response distribution of each V4 population is shown in the same color as in (A). }
\label{fig:ModelSchematic}
\end{figure}

Our model is composed of two layers, a V4 layer and a PFC layer (Fig.~\ref{fig:ModelSchematic}A). We designate the higher cortical area as PFC based on the experimental evidence indicating feedback from PFC as a likely precursor of the delayed responses in V4 \cite{Pasupathy15} (see Results). 
\Hcomment{Removed}\Ecomment{Given how specific we are about modeling PFC / v4 in the above / below, seems to me we could remove the comments on IT etc above?}

The V4 layer is composed of three units: two that are selective for each of the two visual shapes that are being discriminated, namely, shape A and shape B  (Fig.~\ref{fig:ModelSchematic}A, V4 unit 1 (green) and V4 unit 2 (blue), respectively), and a third V4 unit that responds selectively to the occluder-specific features, such as color (Fig.~\ref{fig:ModelSchematic}A,V4 unit 3 (red)). Such selectivity for stimulus shape and color has been previously demonstrated in area V4\cite{Pasupathy99, Bushnell11}.  Each V4 unit can be interpreted as a sub-population of V4 neurons with similar tuning properties. 

%%%%%%%%%%%%%%%%%%%%%%%%%%%%%%%%%%%%%%%%%%%%%%%%%%%
%%%%%%%%%%%%%%%%%%%%%%%%%%%%%%%%%%%%%%%%%%%%%%%%%%%

The model includes two PFC units, which represent two distinct neuronal populations in PFC. While the roles of PFC neurons are not well-understood, PFC is believed to be involved in planning complex behavior and tasks involving short-term memory \cite{Miller01}. Experimental recordings\cite{Pasupathy15} from PFC also show that a subset of PFC neurons have mild shape selectivity, while also responding strongly to occluders.

The sum of PFC activities weighted by the connection weights between V4 and PFC units (Fig.~\ref{fig:ModelSchematic}A) is represented as the feedback signal to V4 units. The initial connection weights between V4 units and PFC units are chosen so that the PFC units show appropriate selectivity after training. Namely, one of the PFC units in the model is designated to be weakly shape A-selective and the other PFC unit is weakly shape B-selective. Both PFC units respond strongly to partially occluded shapes, and only weakly to unoccluded shapes.  \Acomment{more appropriate for results?}  \Ecomment{I agree, a good point here about results for as a home for this ...We don't really put in these response properties of PFC, right, they emerge from training weights etc ... so I can see stating that as a result!}\Hcomment{I added the last paragraph. The initial weights were chosen so that such selectivity emerges. Also this may refer to Eric's next comment. Alternatively, we can cut out this whole paragraph. The initial conditions for the weights are mentioned in more detail later in the training protocol section. }

In this way, PFC neurons of the model respond strongly to both the task-relevant visual features (shape identity) and nuisance variables (occlusion level), while each of V4 populations responds preferentially to single feature of the input visual stimulus. Thus, although the V4 responses are already modulated by both shape and occlusion level, the signals become even more mixed as they go up in the hierarchy. Previous studies have shown that such mixed selectivity in the PFC plays an important computational role in a high-dimensional population encoding of task-relevant information \cite{Rigotti13,Fusi16}. 

\subsection*{Probabilistic network model}

As we detail further below, the responses of the neuronal units evolve toward values that maximize the posterior probability of these responses given the input shape stimulus. In other words, the neuronal activities, and synaptic weights at a slower time scale\Hcomment{Now I have an additional paragraph in Model Architecture which talks about the syn weights}\Ecomment{Is this the first time we've referred to the syn weights?  Not sure but let's make sure we refer to them / define what they mean more generally before talking about how they evolve}, are found by estimating the most likely values given the shape stimulus. 

In our model, visual inputs are simplified and represented by $\kappa$, which includes the shape identity $s$ (shape A or shape B, $s\in\{A,B\}$) and the degree of occlusion $c$ ($c\in[0,1]$), so that $\kappa=(s,c)$. We assume that the V4-PFC circuitry builds a two-level hierarchical description of the input stimulus $\kappa$, via firing rates of V4 ($\mathbf{r_{v4}}$) and PFC neurons ($\mathbf{r_{pfc}}$).\Hcomment{Revised} \Ecomment{Feels redundant in here -- if can refer to above and compress, would improve ...} As it is assumed that each successive random variable is conditionally dependent only on the random variable in the adjacent higher level, the posterior probability of the V4 and PFC responses given $\kappa$ factors as

\begin{align}\label{eq:posterior_full} 
\begin{split}
p(\mathbf{r_{v4}},\mathbf{r_{pfc}}|\kappa) 
& = h_0\cdot p(\kappa | \mathbf{r_{v4}},\mathbf{r_{pfc}}) p(\mathbf{r_{v4}},\mathbf{r_{pfc}})\\
& =  h_0\cdot p(\kappa |\mathbf{r_{v4}},\mathbf{r_{pfc}}) p(\mathbf{r_{v4}} | \mathbf{r_{pfc}}) p(\mathbf{r_{pfc}})\\
&= h_0\cdot p(\kappa|\mathbf{r_{v4}})p(\mathbf{r_{v4}}|\mathbf{r_{pfc}}) p(\mathbf{r_{pfc}})\\
&= h\cdot p(\kappa|\mathbf{r_{v4}})p(\mathbf{r_{v4}}|\mathbf{r_{pfc}}),
\end{split}
\end{align}

\noindent 
where $h_0$ and $h$ are constants. The first equality comes from Bayes' theorem, with a normalization term $h_0$. The second equality is simply a property of joint probability. The third equality is based on the assumption that the probability distribution is set up hierarchically.\Hcomment{I think this can be removed.}\Ecomment{same comment on redundancy -- seems sentences around here can be compressed; also do we need to introduce spatially markov or can we remove that phrase} Based on the assumption of spatially Markovian inference \cite{Lee03,Rao99a,Friston09a,Bogacz15}, \Ecomment{isn't this the same as the conditional independence assumption above eqn 1?  if so, refer back to that ... and if want to use the Markovian terminology ... should introduce that first there (and refs) so reader doesn't think it's something new when we get to this point, rightj?} we made a simplification $p(\kappa|\mathbf{r_{v4},r_{pfc}}) = p(\kappa|\mathbf{r_{v4}})$ in Eq. \ref{eq:posterior_full}.  
Finally, a flat prior on the PFC firing rates is assumed, which is embedded in the constant $h$ on the last line of Eq. \ref{eq:posterior_full}, and therefore, the posterior probability of the neuronal responses is

\begin{align}\label{eq:posterior} 
\begin{split}
p(\mathbf{r_{v4}},\mathbf{r_{pfc}}|\kappa) 
&= h\cdot p(\kappa|\mathbf{r_{v4}})p(\mathbf{r_{v4}}|\mathbf{r_{pfc}}).
\end{split}
\end{align}
\Hcomment{Removed}  \Ecomment{do you need that last sent. right here or could you remove it as you talk about adjusting/evolving below}

\noindent The firing rates of the V4 and PFC units are given as

\begin{align}
&\bf{r_{v4}}=\begin{bmatrix}
r_{v4,1}\\
r_{v4,2}\\
r_{v4,3}
\end{bmatrix}, \,\,
\bf{r_{pfc}}=\begin{bmatrix}
r_{pfc,1}\\
r_{pfc,2}
\end{bmatrix},
\end{align}
where $r_{v4,1}$ and  $r_{v4,2}$ represent the average firing rates of the shape-selective V4 neuronal populations (preferring shape A and shape B, respectively), and  $r_{v4,3}$ is the average firing rate of the occluder feature-selective V4 population.\Hcomment{Removed} \Ecomment{can you remove that last sentence as specify gaussian below also}

We first describe $p(\kappa|\mathbf{r_{v4}})$ and how V4 firing rates depend on the input stimulus $\kappa$. We define $\boldsymbol\mu$ as the bottom-up representation of the stimulus
\begin{align}
\boldsymbol\mu=\begin{bmatrix}
\mu_{1}\\
\mu_{2}\\
\mu_{3}\\
\end{bmatrix}.
\end{align}
The difference between this bottom-up representation and the V4 responses $\mathbf{r_{v4}}$ gives the noise term $\boldsymbol\eta_{1}$,

\begin{align}\label{eq:noise1} 
\boldsymbol\eta_{1} = \boldsymbol\mu-\mathbf{r_{v4}}, 
\end{align}
which has a Gaussian distribution with zero mean and diagonal covariance matrix 
\begin{align}\label{eq:variance1} 
\Sigma_1 = \begin{bmatrix}
{\sigma_{1}^2} & 0 & 0\\
0 & {\sigma_{2}^2} & 0\\
0 & 0 & {\sigma_{3}^2}
\end{bmatrix}.
\end{align}
The distribution $p(\kappa|\mathbf{r_{v4}})$ is the likelihood of the V4 neuronal activities given the sensory input $\kappa$. Assuming a flat prior on $\mathbf{r_{v4}}$,  $p(\kappa|\mathbf{r_{v4}})\propto p(\mathbf{r_{v4}|\kappa)}$. Thus, 
\begin{align}\label{eq:feedforward_dist} 
p(\mathbf{r_{v4}}|\kappa) = N( \mathbf{r_{v4}}; \boldsymbol\mu,\, \Sigma_1).
\end{align}

The mean $\boldsymbol\mu$ and the covariance matrix $\Sigma_1$ are determined by the input shape identity $s$ and the occlusion level $c$.  Changes in $\boldsymbol\mu$ and $\Sigma_1$ describe the sensory-input driven responses of the V4 populations to different shapes under various degrees of occlusion. In other words, for each occlusion level and the shape identity, there is a most-likely firing rate of each V4 unit given by $\boldsymbol\mu$, and that likelihood falls off according to the covariance $\Sigma_1$.  

\Hcomment{Done}\Ecomment{need topic sentence here} Here we describe how we modulate $\boldsymbol\mu$ and $\Sigma_1$ based on the sensory input $\kappa$. Let's assume the animal is presented with shape A as the test shape. With shape A presented,  $\mu_{1}$, the Gaussian mean of the firing rate distribution of V4 unit 1 in Fig.~\ref{fig:ModelSchematic}A (the shape A-selective V4 population), decreases as occlusion $c$ increases (Fig.~\ref{fig:ModelSchematic}C, green).  On the other hand, $\mu_{2}$ of the V4 population preferring shape B (V4 unit 2 in Fig.~\ref{fig:ModelSchematic}A) stays constant at a ``baseline" firing rate, independent of the change in occlusion level.\Hcomment{Yeah, I think we can remove this.}\Ecomment{can we remove the prev sentence?  seems maybe if we are not simulating the no stimulus case ever?  does that make sense?} That is, the V4 unit 2 does not prefer shape A, it responds with a low firing rate regardless of the occlusion level (Fig.~\ref{fig:ModelSchematic}C, blue).  The standard deviation $\sigma_{1}$ of the preferred V4 unit increases as occlusion increases, in order to capture the increasing uncertainty of the shape identity under higher degrees of occlusion (see Fig.~\ref{fig:ModelSchematic}C, where the green distribution widens as occlusion increases). The standard deviation $\sigma_{2}$ of the non-preferred V4 population (V4 unit 2) is assumed to be constant.  On the other hand, $\mu_{3}$ of the occluder-selective V4 population (V4 unit 3) increases as occlusion level increases, but its $\sigma_{3}$ stays constant, as this population of V4 neurons is not selective for the shape identity (Fig.~\ref{fig:ModelSchematic}C, red). 

The dependence of the means and the variances on the occlusion level $c$ was set to be linear:   $\boldsymbol\mu =  \boldsymbol\mu_0 +\boldsymbol \alpha\cdot c$ and $ \Sigma_1 =  \Sigma_0 + \boldsymbol\beta\cdot c$ with $\boldsymbol\mu_0 = [50\; 20\; 20]^T$, $\boldsymbol\alpha = [-5\; 0\; 100]^T$, $\Sigma_0 =  \mathbf{I_3}$,  and $\boldsymbol\beta = [5\; 0\; 0]^T$. The  slopes ($\boldsymbol\alpha$, $\boldsymbol\beta$) and the values defining the response distributions when the shape is unoccluded ($\boldsymbol\mu_0$, $\Sigma_0$) at $c=0$ , were manually chosen to match the peak firing rates observed in experiments. With this choice of $\boldsymbol\alpha$, the peak of the response distribution decreases, stays constant, and increases for V4 unit 1, 2, and 3, respectively. The values chosen for $\boldsymbol\beta$, on the other hand, indicate that ambiguity of the stimulus feature increases only for the test shape preferred V4 unit 1. In this way, the input stimuli-- shape A and shape B with various degrees of occlusion -- are represented by the response distributions of three different V4 populations given $\kappa$, rather than by using actual pixel images. 

The second term on the right side of Eq.\ref {eq:posterior}, $p(\mathbf{r_{v4}}|\mathbf{r_{pfc}})$,  provides the top-down effects on the posterior distribution, also described as Gaussian. Here, the mean is the prediction made by PFC, $\mathbf{u}\cdot\mathbf{r_{pfc}}$, which is the sum of the two PFC population responses weighted by the connection weight matrix $\mathbf{u}$. In more general cases, this weighted sum is filtered by a nonlinearity $f$, thus yielding the top-down prediction $f(\mathbf{u}\cdot\mathbf{r_{pfc}})$ (Fig.~\ref{fig:ModelSchematic}B). For the simulations in this study, however, the nonlinearity on weighted PFC responses was ignored and the predictions were assumed to be linear, i.e., $ f(\mathbf{u}\cdot\mathbf{r_{pfc}})=\mathbf{u}\cdot\mathbf{r_{pfc}}$, as in~\cite{Rao99a}. 
% Multivariate Gaussian, doing learning in a layer-specific way, marginal distribution remains Gaussian.
The connection weights between the V4 and PFC neuronal units are given as

\begin{align}
\bf{u}=\begin{bmatrix}
u_{1,1} & u_{1,2} \\
u_{2,1} & u_{2,2} \\
u_{3,1} & u_{3,2} \\
\end{bmatrix}.
\end{align}

The difference between $\mathbf{u}\cdot\mathbf{r_{pfc}}$, the top-down prediction made by PFC, and the V4 responses $\mathbf{r_{v4}}$ is then

\begin{align}\label{eq:noise2} 
\boldsymbol\eta_{2} = \mathbf{r_{v4}}-\mathbf{u}\cdot\mathbf{r_{pfc}},
\end{align}
where the noise $\boldsymbol\eta_{2}$ has a Gaussian distribution with zero mean and covariance matrix $\Sigma_2$,

\begin{align}\label{eq:variance2} 
\Sigma_2 = \begin{bmatrix}
{\sigma'}_{1}^2 & 0 & 0\\
0 & {\sigma'}_{2}^2 & 0\\
0 & 0 & {\sigma'}_{3}^2
\end{bmatrix}.
\end{align}

The distribution of V4 responses given the PFC responses, $p(\mathbf{r_{v4}}|\mathbf{r_{pfc}})$, is then 

\begin{align}\label{eq:feedback_dist}
p(\mathbf{r_{v4}}|\mathbf{r_{pfc}}) = N( \mathbf{r_{pfc}}; \mathbf{u}\cdot\mathbf{r_{pfc}},\,\Sigma_2).
\end{align}

The standard deviation of the response distribution of each V4 unit given the PFC responses determines the relative significance of the top-down predictive contribution on shaping the V4 responses. Specifically, a smaller standard deviation leads to smaller noise terms, forcing closer matches between PFC and V4 responses. These standard deviations were chosen as ${\sigma'}_{1} = 10,\, {\sigma'}_{2} = 10$, and  ${\sigma'}_{3} = 1$. Thus, the top-down component is more strongly emphasized for V4 unit 3,  the V4 neuronal population selective for  occluders. We found that such emphasis on the predictive component for the occluder-selective V4 population was necessary to reproduce the experimentally observed PFC response characteristics -- an increase in PFC responses with a rise in occlusion level (see Results).

Given the visual stimulus $\kappa$, the firing rates $\mathbf{r_{v4}}$ and $\mathbf{r_{pfc}}$ adjust in order to maximize the posterior distribution, namely, $p(\kappa|\mathbf{r_{v4}}) p(\mathbf{r_{v4}}|\mathbf{r_{pfc}})$. Maximizing this is equivalent to minimizing its negative logarithm, which is defined as the cost function $E$,

\begin{align}\label{eq:cost} 
\begin{split}
E = 
&\left(\mathbf{r_{v4}}- \boldsymbol\mu\right)^T\Sigma_1^{-1}\left(\mathbf{r_{v4}}- \boldsymbol\mu\right)\\
&+\left(\mathbf{r_{v4}}-\bf{u}\cdot\mathbf{r_{pfc}}\right)^T \Sigma_2^{-1} \left(\mathbf{r_{v4}}-\mathbf{u}\cdot\mathbf{r_{pfc}}\right).
\end{split}
\end{align}

Note that this cost function is the sum of the squared error $\boldsymbol\eta_{1}^T\boldsymbol\eta_{1}$ between the V4 responses and the sensory-input imposed representation, and the squared error $\boldsymbol\eta_{2}^T\boldsymbol\eta_{2}$ between the V4 responses and the top-down prediction made by PFC, weighted by their inverse variances. 

The optimal ``parameters'' -- the neuronal responses and the connection weights --\Ecommentnew{in most places where -- used need to adjust spacing around to make it even, here missing a space after weights} are thus found by minimizing this cost function $E$ with respect to the parameters $\mathbf{r_{v4}}$, $\mathbf{r_{pfc}}$, and $\mathbf{u}$. The initial V4 responses in experiments, that presumably depend only on the feedforward sensory input, are found by minimizing only the first term of Eq.\ref{eq:cost}. The initial responses are therefore equal to the sensory-driven representation $\boldsymbol\mu$. However, the delayed V4 responses, which we hypothesize to depend on both the feedfoward sensory input and the feedback prediction, are found by minimizing the entire cost function Eq.\ref{eq:cost}.

\subsection*{Training protocol: weight adjustment during the preliminary phase}

We divide the optimization process into two phases based on the experimental setup: the preliminary phase and the test phase. In this section, we discuss how the synaptic weight matrix between PFC and V4 \Hcomment{Corrected}\Ecomment{should we say between PFC and V4 here and below?} is found during the preliminary phase. To find these weights, we minimized the cost function $E$ with respect to $\mathbf{r_{v4}}$ and $\mathbf{r_{pfc}}$ as well as with respect to the connection weight matrix $\mathbf{u}$, over a series unoccluded trials. Then during the test phase, the optimal estimates of the neuronal responses to shapes under varying degrees of occlusion are determined by minimizing the cost function with respect to $\mathbf{r_{v4}}$ and $\mathbf{r_{pfc}}$, with the connection weights fixed at the learned values. 

The preliminary phase corresponds to the stage at the beginning of the experiment where the animal is exposed to a pair of unoccluded shapes used for the experimental session. During this period, the animal is presented with a selected pair of unoccluded shape stimuli several times ($\sim 20$), while performing the matching task.  We introduced its equivalent in the simulation, during which the cost function $E$ is minimized by gradient descent with respect to the firing rates of the V4 units $\mathbf{r_{v4}}$ and PFC units $\mathbf{r_{pfc}}$, as well as the connection weight matrix  $\mathbf{u}$. During this phase, unoccluded shape A and shape B are randomly chosen and used as inputs to the model for up to 30 trials. Thus, over this phase the synaptic weight matrix is learned over the course of these multiple trials with unoccluded shapes.

The optimal estimates of $\mathbf{r_{v4}}$, $\mathbf{r_{pfc}}$, and $\mathbf{u}$ are obtained by performing gradient descent on $E$ with respect to these parameters at different learning rates:

\begin{align}\label{eq:descent}
\begin{split}
&\frac{d\mathbf{r_{v4}}}{dt} = -k_r \frac{\partial E}{\partial \mathbf{r_{v4}}}\\
&\frac{d\mathbf{r_{pfc}}}{dt} = -k_r \frac{\partial E}{\partial \mathbf{r_{pfc}}}\\
&\frac{d\mathbf{u}}{dt} = -k_u \frac{\partial E}{\partial \mathbf{u}}.
\end{split}
\end{align}

%\noindent We perform gradient descent with respect to all the parameters in the network. In hierarchical Bayesian models with more than two layers, alternatively, iterative methods such as particle filtering can be used until the variables across all levels are stabilized \cite{Lee03}

The learning rate of $\bf{u}$ was a significantly smaller value $k_u = 0.001$, compared to that of $\mathbf{r_{v4}}$ and $\mathbf{r_{pfc}}$, which was  $k_r = 0.1$.  This models the relatively faster dynamics of firing rates and slower dynamics of synaptic plasticity. For each selected shape, we carried out gradient descent  either until the firing rates reach steady states after a minimum 20 iterations, or until the iteration exceeds the maximum of 500 iterations.  While $\mathbf{r_{v4}}$ and $\mathbf{r_{pfc}}$ rapidly converge to a fixed point for each of the sampled shapes, the connection matrix $\mathbf{u}$ gradually converges over the course of multiple samples of shape A and B. In this way, the weight matrix $\mathbf{u}$ is tuned over the course of the preliminary phase, which corresponds to the animal's familiarization with the pair of the shapes at the beginning of the experiment. We set initial weights for $u_{1,2}$ and $u_{2,1}$ smaller than the initial values of other connection weights, to slightly bias one of the PFC populations (PFC unit 1) to be shape A-selective and the other (PFC unit 2) to be shape B-selective. %With these initial conditions, weakly shape-selective PFC populations emerge (see Results). Specifically, the learned weight matrix has positive $u_{1,1}$ and $u_{2,2}$ that have larger amplitudes than $u_{1,2}$ and $u_{2,1}$ which are also positive, so that one of the PFC units is more selective for shape A and the other PFC unit is more selective for shape B. 
\Ecommentnew{in above, should flag forward to results where develop this, it does read like a result ... or shorten if possible}

We acknowledge a limitation of the gradient descent method on $E$ in Eq.\ref{eq:descent}, which is that requires nonlocal  computation. In other words, the activities and the synaptic strengths of all the neuronal units in the system must  be known in order to take a gradient descent step, a requirement that is not physiologically realistic. This issue also exists in previous models of predictive coding and sparse coding in the visual system \cite{Rao99a,Olshausen96,Olshausen97}, as pointed by \cite{Bogacz15,Zylberberg11}. While we do not pursue this matter further here, we note that Zylberberg et al \cite{Zylberberg11} shows that in the limit that the neuronal activity is sparse and uncorrelated, the non-local gradient descent rule is approximately equivalent to a synaptically local rule.

\subsection*{Optimal stimulus representation during the test phase}

Once the weight matrix $\mathbf{u}$ has converged over the course of the preliminary phase, it is fixed at the learned values during the test phase. The test phase corresponds to the recording session where the animal performs the matching task while test shapes with varying degrees of occlusion are displayed.  We hypothesize that the V4 and PFC recordings from the experiment are represented by the average firing rates of the V4 and PFC populations in the model network, $\mathbf{r_{v4}}$ and $\mathbf{r_{pfc}}$, that minimize the cost function $E$. Either shape A or shape B can be used as the input to the network. In this paper, however, without loss of generality we only show the simulations with shape A as the test shape so that the V4 unit selective for shape A (V4 unit 1) is the ``preferred" population and the shape B-selective V4 unit (V4 unit 2) is the ``non-preferred" population.   The weight matrix $\mathbf{u}$ is fixed at the learned values from the preliminary phase. 

For each occlusion level, the optimization is carried out in two parts, to reflect the dynamics of the V4 responses. The initial responses of V4 neurons observed in experiments are compared to the V4 responses $\mathbf{r_{v4}}$ that minimize the first part of the cost function $E$ (Eq.\ref{eq:cost}), namely, 

\begin{align}\label{eq:cost1} 
\begin{split}
E_1 = 
&\left(\mathbf{r_{v4}}- \boldsymbol\mu\right)^T\Sigma_1^{-1}\left(\mathbf{r_{v4}}- \boldsymbol\mu\right).
\end{split}
\end{align}
$E_1$ is simply a weighted difference between the V4 neuronal responses and the V4 responses predicted by the bottom-up sensory input. Therefore, $\mathbf{r_{v4}}$ that minimizes $E_1$ are interpreted as the V4 responses shaped by only the feedforward inputs.  \Hcomment{The last paragraph in the previous section talks about weight training and emergence of PFC tuning...?}\Ecomment{this feels redundant to text in last para above Training Protocol: section above.  Not sure if there is any good way to adjust}

On the other hand, the delayed responses of V4 neurons, as well as the PFC responses, are found by minimizing the entire cost function $E$ (Eq.\ref{eq:cost}) with respect to $\mathbf{r_{v4}}$ and $\mathbf{r_{pfc}}$. We rewrite the full cost function $E$ as $E_2$: 

\begin{align}\label{eq:cost2} 
\begin{split}
E_2 = E = 
&\left(\mathbf{r_{v4}}- \boldsymbol\mu\right)^T\Sigma_1^{-1}\left(\mathbf{r_{v4}}- \boldsymbol\mu\right)+\left(\mathbf{r_{v4}}-\bf{u}\cdot\mathbf{r_{pfc}}\right)^T \Sigma_2^{-1} \left(\mathbf{r_{v4}}-\mathbf{u}\cdot\mathbf{r_{pfc}}\right).
\end{split}
\end{align}

$E_2$ includes a term that depends on the difference between $\mathbf{r_{v4}}$ and the top-down predictions made by PFC, $\mathbf{u}\cdot\mathbf{r_{pfc}}$, in addition to the error term between the $\mathbf{r_{v4}}$ and the V4 responses predicted by the input visual stimulus. Therefore, $\mathbf{r_{v4}}$ that minimizes this cost function $E_2$ is interpreted as the V4 responses shaped by both the feedforward and the feedback signals. This $\mathbf{r_{v4}}$ is compared to the delayed responses in V4 neurons in experiments that we hypothesize  to be induced by feedback from PFC.

$E_1$ and $E_2$ are minimized using gradient descent and MATLAB \textit{fminsearch} with respect to $\mathbf{r_{v4}}$ and $\mathbf{r_{pfc}}$, starting from the initial value at 10 (spikes/s) for all neuronal units. The average responses of each neuronal unit thus found are compared to experimentally measured neuronal responses to the shape stimuli with varying degrees of occlusion (Fig.~\ref{fig:Experiment}, \ref{fig:Result_Main}).

\Acomment{I just skimmed through most of the methods. Did not read it thoroughly. I figure, you and Eric will make sure all the equations are correct and appropriately described. Two broad comments: It may be best to start with Model architecture and then go to assumptions so you provide a framework before explaining assumptions. \Hcomment{Now I removed the second assumption from Coding Assumptions, and think it flows better}\Hcomment{Hmm... this is where everyone's opinion diverges-I've changed this order a few times. I will put it aside for now and get back to this} Second, Try to avoid repeating things -- for example, minimizing with grad desc comes up several times. Might be good to tighten those up\Hcomment{Yes, did some revisions on this }}

\section*{Results}

We first present experimental evidence that supports the hypothesis that feedback signals from PFC modulate shape representations in V4 (\textit{Experimental evidence for feedback signals in area V4}). We then compare the outcomes in our probabilistic network model (\textit{Structure and design of probabilistic network model}) to physiology and explain how robust shape recognition can be achieved in our model (\textit{Neuronal dynamics predicted by hierarchical Bayesian inference}). Subsequently, we identify necessary assumptions on the network structure (\textit{Parsimony of the network structure}) and the signal structure (\textit{Implication of inverse variance: differential weighting of feedforward and feedback inputs}) of the model to capture the key trends in the experimental results. Finally, using our model, we make predictions on shape selective neuronal responses to  a new type of reduced stimulus clarity (\textit{Model prediction on shapes obscured by non-salient occlusion, noise, or reduced contrast}). \Acomment{I edited the above para}

\subsection*{Experimental evidence for feedback signals in area V4}

\begin{figure}[h]
\noindent \centering
\includegraphics[scale=0.7]{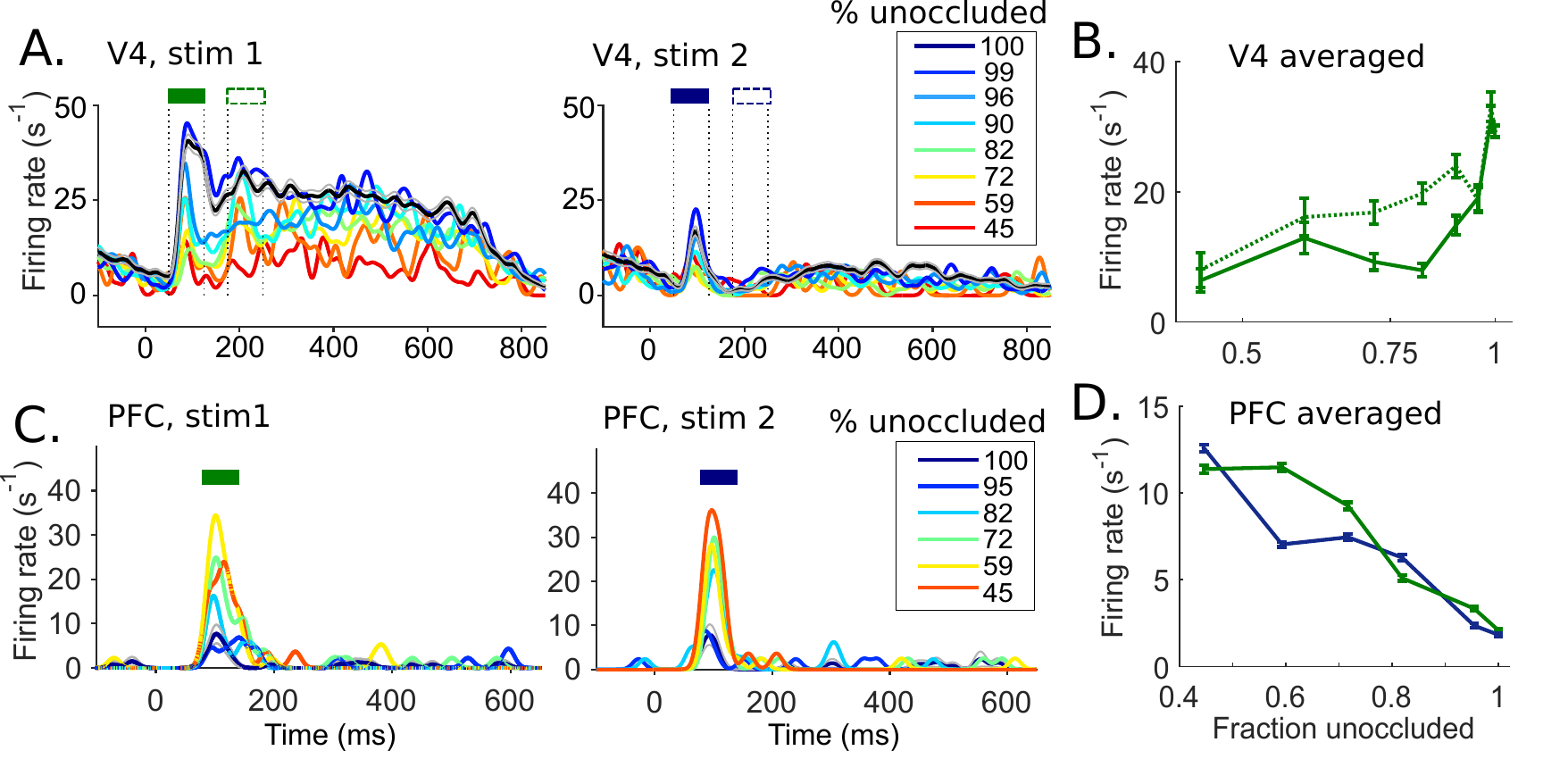}%{Experiment}
\caption{{\bf Recordings from V4 and PFC show characteristic response dynamics.}
(A) Example V4 cell responses to a preferred (left) and a non-preferred shape (right) during the discrimination task. Test stimulus onset was at time 0 ms. Level of occlusion was measured by \% unoccluded area (line color). Black line (100\% unoccluded) represents the unoccluded stimulus. Two transient peaks are identified by filled and open rectangles. (B) The time averaged V4 firing rates during the initial and the delayed peaks (identified in A) as a function of occlusion level. Solid lines show averaged firing rates for the preferred shape during the initial peak, and the dotted lines indicate averaged firing rates during the delayed transients, as marked above response traces in (A).  (C) Response of an example PFC cell to the two shape stimuli (left and right) during the discrimination task. (D) Averaged PFC responses as a function of occlusion level.  Responses to each of the two shapes are shown in green and blue, respectively. Population data follow the same trend. Data adapted with permission from \cite{Pasupathy15}.} 
\label{fig:Experiment}
\end{figure}

Recent experiments demonstrated that neurons in V4 and PFC show strikingly different response patterns in monkeys performing a sequential shape discrimination task. %\Hcheck{In the experiments, animals report whether two stimuli presented sequentially are the same or different, where the second stimulus is partially occluded. Note that the animals were exposed to the pair of unoccluded stimuli at the beginning of each session. Responses from V4 and PFC were recorded while the animals were presented with the test shape under varying degrees of occlusion. The occlusion level was titrated by randomly positioning occluding dots with the size systematically varied, and quantified by the fraction of the stimulus area that was not occluded. The stimuli were presented in a color preferred by the cell while the occluding dots were in a non-preferred color. The recordings from both areas were made separately, and data from both correct and error trials were included in the analysis.} 
Fig.~\ref{fig:Experiment}A shows the response dynamics of an example V4 cell to a preferred shape (left) and a non-preferred shape (right). The V4 neuron exhibits two transient peaks when the preferred shape was presented, but  only one smaller peak for the non-preferred shape.  In the initial transient at the onset of the preferred shape stimulus, the V4 neuron responded strongly to the unoccluded shape (black), and an increase in occlusion weakened the shape selective responses (color). While the first peak shows a dramatic dependence on occlusion, the latter peak of responses shows a weaker dependence. Fig.~\ref{fig:Experiment}B shows the averaged responses of the V4 neuron during the initial transient (50-125 ms) and the delayed transient (175-250 ms), illustrating the differential effects of occlusion on V4 responses over time.  The reduced effect of occlusion on V4 responses to the preferred shape during the second transient leads to enhanced shape selectivity, as previously observed in \cite{Pasupathy15, Kosai14}.

%%%%%%%%%%%%%%%%%%%%%%%%%%%%%%%%%%%%%%%%%%%%%%%%%%%
%%%%%%%%%%%%%%%%%%%%%%%%%%%%%%%%%%%%%%%%%%%%%%%%%%%

In contrast to V4 neurons, PFC neurons exhibit one peak, and show strongest responses to occluded stimuli and weakest responses to unoccluded stimuli, as shown for an example PFC neuron in Fig.~\ref{fig:Experiment}C \cite{Pasupathy15}. Fig.~\ref{fig:Experiment}D shows the time averaged responses of the PFC neuron as a function of occlusion level, for both the preferred and the non-preferred shapes. As occlusion increases, the PFC responses increase, which is the opposite trend as for V4.  Moreover, the timing of the peak PFC responses is between the initial and the delayed transients of V4 responses, consistent with the hypothesis that the PFC responses, which arise from  feedforward transmission of sensory information, in turn send feedback inputs and drive the second peak of responses in V4. These experimental observations led us to the hypothesis that the feedback inputs from PFC and other higher cortices underlie delayed improvement of shape selective responses under occlusion in V4. For more details on the experimental results, see \cite{Kosai14} and \cite{Pasupathy15}.

\subsection*{Structure and design of probabilistic network model}

We sought to understand the response dynamics of V4 and PFC neurons in the context of predictive coding, a hierarchical encoding of stimuli widely used to probe interactions of lower and higher sensory areas. We first pose a probabilistic network model of the V4-PFC circuitry with the presumptive feedback based on predictive coding, and introduce an innovation that differentiates our model from previous predictive coding models. 

In each layer of our V4-PFC network model, there are distinct units, each of which represents a neuronal population with similar tuning properties. The V4 layer is composed of three units which respond preferentially to different features of the visual stimulus (Fig.~\ref{fig:ModelSchematic}A): unit 1 to shape A, unit 2 to shape B, and unit 3 to an occluder-specific feature, for example the color of the occluders. In PFC, there are two units that respond strongly to occlusion, while also exhibiting some degree of shape selectivity. The shape-selective V4 units and the PFC units were motivated directly by physiology, and the occluder-selective V4 unit was included in order to capture the response patterns in the experiments, as explained in more detail later. The representation of a population of neurons as a single unit is a common simplification but we find that each unit replaced by a population of multiple neurons with mild heterogeneity yields qualitatively the same response trends as with the single unit model (See S1 Text, S1 Fig). 
 
In the model, V4 receives feedforward sensory inputs and seeks to match the responses imposed by the sensory inputs. At the same time, feedback predictions from PFC bias the V4 responses. The weighted sums of PFC responses provide top-down predictions conditioned on underlying visual stimulus, and are regarded as the feedback from PFC to V4. These predictions are compared to the initial \Ecomment{word initial inserted, please check it's correct} V4 neuronal responses, as the system attempts to minimize the difference between the top-down predictions and the V4 responses\Ecomment{should the last few words be something like ... and the feedforward sensory input or something like that?  Except that I guess that would be too redundant with first sentence in para that follows ...}.  \Ecomment{**}\Hcomment{Paragraphs revised} 

This process is equivalent to finding the most likely neuronal responses given the visual stimulus. With hierarchical Bayesian inference assumed, the most likely representation of the responses is obtained by finding a set of responses that maximize the posterior probability given the visual stimulus, which is equivalent to the product of conditional probabilities of the neuronal activities given only the activities of the next higher area (See Materials and Methods, Eq.\ref{eq:posterior_full}). Here, the visual input to each V4 unit is represented as a Gaussian distribution, whose mean and variance change according to the shape identity and the occlusion level (Fig.~\ref{fig:ModelSchematic}C). Similarly, the feedback from PFC to each V4 unit is  described by a Gaussian distribution with the peak at a sum of the PFC responses weighted by the synaptic strengths (Fig.~\ref{fig:ModelSchematic}B).    
\Ecomment{Can you work the material in this paragraph at the end of the paragraph marked ** above?  seems would flow better?  Maybe the same is true for the paragraph that follows?  In this way, you end the section with the ``result' ' which is your reformulation?} \Hcomment{Paragraphs revised}

In this way, the optimal representation of the neuronal responses integrates both the bottom-up sensory input and the top-down prediction.  This is done by minimizing a cost function composed of the difference between the V4 activities and the top-down predictions as well as the difference between the V4 activities and the V4 responses predicted by the sensory input, with each term inversely weighted by its respective variance (See Materials and Methods, Eq.\ref{eq:cost}). We compare this optimal representation directly to the neuronal responses in experiments; this differs from previous studies \cite{Rao99a,Srinivasan82} where the residual error between the prediction and the neuronal activity was associated with physiologically measured responses. With this reformulation, neural activity conveys both the sensory input and the internal prediction, preventing the situation \Hcomment{preventing instead of avoiding?}\Ecomment{better word there?  } in original implementations of predictive coding in which neurons have depressed activity when familiar stimuli are presented. 

\subsection*{Neuronal dynamics predicted by predicting coding model}

In this section, we show that the proposed probabilistic network model (Fig.~\ref{fig:ModelSchematic}), with the synaptic weights trained as in the experiment, captures the key physiological properties observed in the V4 and PFC response dynamics.

\subsubsection*{Network training and synaptic weight matrix}

First, the network was trained following the experimental procedure where the animal was exposed to the pair of unoccluded shapes. During this preliminary phase, the connection weight matrix $\bf{u}$ between PFC and V4 is learned by gradient descent on the cost function $E$ with respect to the weights $\bf{u}$ as well as the neuronal responses $\bf{r_{v4}}$ and $\bf{r_{pfc}}$, while unoccluded shape stimuli randomly selected from the set of shape A and shape B, are input to the network. The learning rate for neuronal firing rates is significantly larger than that for weights (See Materials and Methods, Eq.~\ref{eq:descent}). \Hcomment{Done} \Ecomment{maybe refer back to equation number in methods where this is laid out}Thus, for each sampled shape, the firing rates of the neuronal units converge rapidly. The weight matrix $\bf{u}$ converges on a slower time scale, over the course of the preliminary phase with multiple presentations of unoccluded shapes. With initial values of the connection weights set to

\begin{align*}
\bf{u}=\begin{bmatrix}
1& -1 \\
-1 & 1\\
1 & 1 \\
\end{bmatrix},
\end{align*}
the connection weight matrix converges to

\begin{align*}
\bf{u}=\begin{bmatrix}
2.32& 0.21 \\
0.26 & 2.37\\
0.94 & 0.94 \\
\end{bmatrix},
\end{align*}
where the asymmetric weights between the PFC units and the shape-selective V4 units indicate shape selectivity in PFC units. The shape selectivity in PFC units and resulting response characteristics are preserved as long as the initial values for $u_{1,2}$ and $u_{2,1}$ are sufficiently smaller than $u_{1,1}$ and $u_{2,2}$ to introduce an initial bias on shape selectivity. 

\Hcomment{Revised the paragraph}\Ecomment{add topic sentence / rearrange a bit?  Almost seems that the sentence starting The convergence of ... sentence below seems like could be a good choice of topic sent for this para?} 
The convergence of the weight matrix depends on the choice of initial conditions, given the non-convex and under-constrained nature of the cost function $E$, as there are multiple combinations of the connection weights and neuronal responses that minimize $E$. However, this does not limit our main results, as we can regard the biased initial values as the connections between a subset of PFC populations and the V4 population of interest before learning the shapes, which may have either weak negative values or positive values, among a wide range of random initial connection weights between PFC and V4. Depending on the initial connection weights, the connections will either become stronger or weaker over the course of training, and shape selectivity in PFC neurons emerges. 

The obtained connection weight matrix is interpreted as a stored template or memory of the shape pair, and is fixed during the following test phase. The memory of the shapes encoded in the connection weights is similar to the idea proposed in \cite{Mumford92} where it was suggested that descending pathways store templates in the weights of their synapses. \Ecommentnew{I'd add the references to mumford etc here, and consider removing from methods}

\subsubsection*{Two-step inference on neuronal activity}

With the trained connection weights, we find the model responses of each unit to partially occluded stimuli are comparable to neuronal responses in experimental recordings during the sequential shape discrimination task described above. In particular, we separate the responses inferred strictly by feedforward sensory inputs from those generated by integrated signals of both feedforward inputs and feedback predictions, and show that the model responses capture the temporal dynamics in the electrophysiological recordings.  \Hcomment{Revised}\Ecomment{State above in more active  / direct way as per Anitha's comment at end of this section ... We show that ... model can capture ... etc}

\begin{figure}[!htb]%[hp]
\noindent \centering
\includegraphics[scale=0.65]{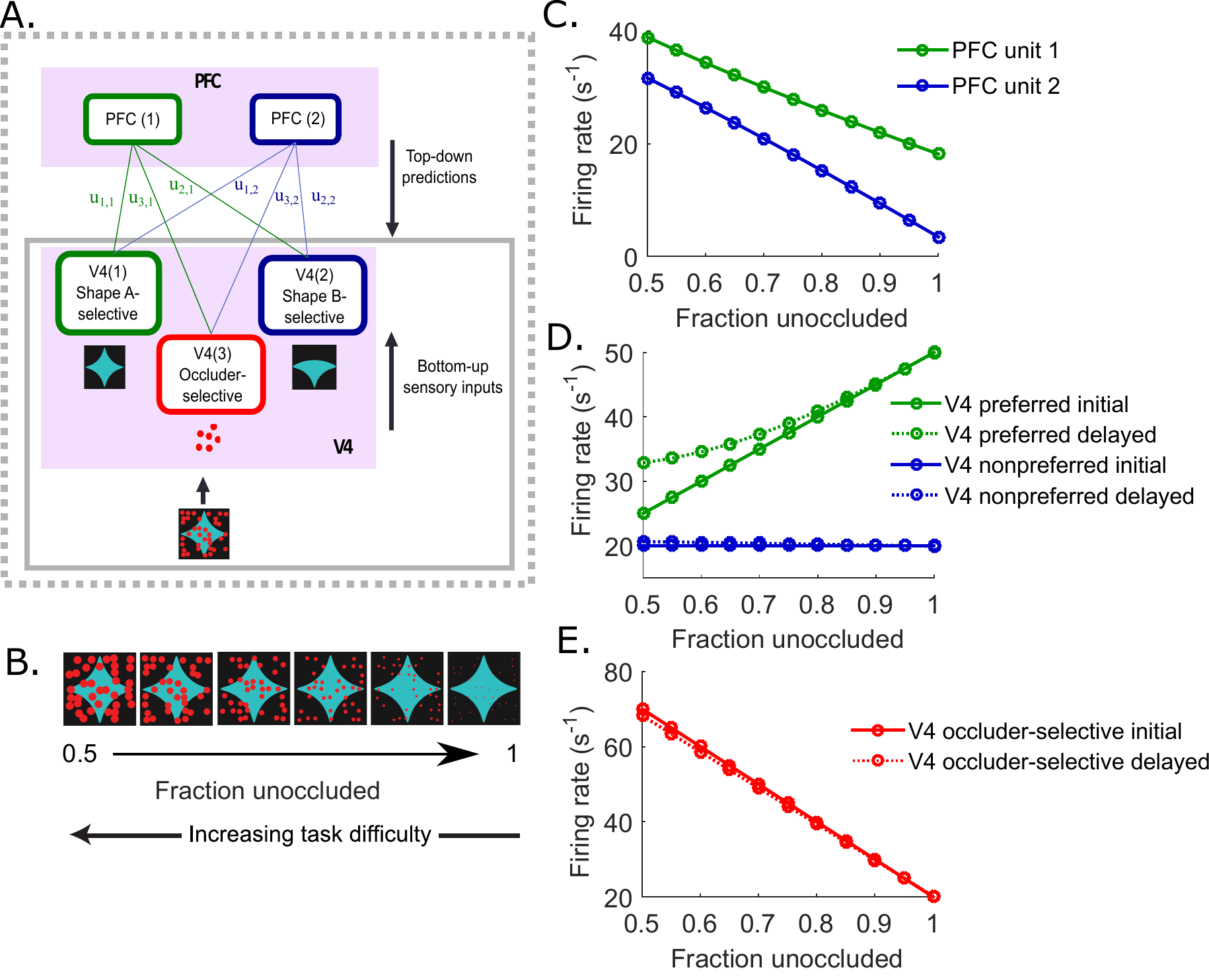}
%includegraphics[scale=0.7]{Result_Main}
\caption{{\bf Model simulations. The optimal representation based on hierarchical Bayesian inference reproduces V4 and PFC responses in the experiments.}
(A) The network model schematic as in Fig.~\ref{fig:ModelSchematic}A. The solid rectangle shows the initial feedforward-only signal computation. The dotted rectangle encompasses the computations for the delayed response inferences that integrate the bottom-up sensory inputs and the top-down predictions from PFC. The corresponding optimal representations are shown in solid (initial, feedforward-only) and dotted (delayed, feedforward+feedback) lines in D-E. (B) Illustration of the input stimuli-- shape A with varying degrees of occlusion. The actual images were not used as the input; the $\kappa$-dependent population response distributions of V4 neurons were used to represent the shape stimuli. Note that the occluders are of a different color than the shape or the background, and activate a group of V4 cells selective for the color.  (C) Inferred PFC responses increase as occlusion level increases, in accordance with experiments. A weak shape selectivity is present, as PFC unit 1 responds at higher rates than PFC unit 2 to the presented shape A across the occlusion levels. (D) Inferred responses of the shape-selective V4 units before (solid) and after (dotted) the top-down prediction. The green lines are the optimal responses of the V4 population selective for the test shape-- shape A (V4 unit 1), and the blue lines are those of the non-preferred V4 population that responds preferentially to shape B (V4 unit 2). 
(E) Model prediction of average firing rates of the occluder-selective V4 population (V4 unit 3), as a function of occlusion level. The salient occlusion activates this class of V4 neurons. Note that the x-axis shows fraction unoccluded.}
\label{fig:Result_Main}
\end{figure}

The optimal representations of the neuronal responses $\bf{r_{v4}}$ and $\bf{r_{pfc}}$ that minimize either the first term ($E_1$ from Eq.\ref{eq:cost1}) or the full representation of the cost function $E$ ($E_2$ from Eq.\ref{eq:cost2}) are computed at each occlusion level. As explained in Materials and Methods, these are equivalent to the optimal responses in hierarchical Bayesian inference that maximize the posterior probability of the V4 neuronal responses given the shape identity and the occlusion level.  Here we assume that the occluders are of a color different from that of the shape or the background, i.e., occlusion is salient and distinct (Fig.~\ref{fig:Result_Main}B). The occluders therefore activate V4 unit 3, the occluder-selective neuronal population in the model.\Hcomment{Moved to Methods} \Ecomment{I'd look for sentences that can be removed to address Anitha's comment ... for example, seems clear that last sentence in this para could definitely be removed or moved to methods ... might be able to omit grad descent detail too?}

%%%%%%%%%%%%%%%%%%%%%%%%%%%%%%%%%%%%%%%%%%%%%%%%%%%
%%%%%%%%%%%%%%%%%%%%%%%%%%%%%%%%%%%%%%%%%%%%%%%%%%%

We make the inference on the neuronal responses in two steps. First, only the bottom-up sensory input is considered, so that the posterior distribution depends only on the stimulus $\kappa$ (Fig.~\ref{fig:Result_Main}A, solid box). In other words, the optimal representations of the activities of the V4 units, $\bf{r_{v4}}$, are found by minimizing only the first term of the cost function $E$ in Eq.\ref{eq:cost}, or equivalently, by maximizing Eq.\ref{eq:feedforward_dist}. We hypothesize that these optimal responses modulated only by the bottom-up sensory inputs, \Hcomment{Revised}\Ecomment{I'd rephrase as active voice ... we xxxx ...  and search for similar uses of passive voice that can rephrase} to correspond to the initial transient in recorded V4 responses. Thus, only feedforward signals are present at this stage.  \Hcomment{Can we discuss more on this? I think it will be clearer to compare the initial and the delayed responses side-by-side within one paragraph... also I kind of like having PFC responses on top next to the model schematic and same for V4, rather than switching the order so that V4 responses are on top next to PFC units in the schematic. I guess it's not critical... but looks nicer?} \Ecomment{I wonder if one way to address Anitha's comment below would be to discuss the Fig 3D and E, initial responses here.  Then move on to delayed.  That might require swapping panels Fig 3 DE above Fig 3C in fig?}

The delayed transients in V4 responses following the peak of responses in PFC, on the other hand, are compared to the optimal responses that integrate both the bottom-up and the top-down inputs. The model representations of the delayed V4 responses and the PFC responses, therefore, are obtained by finding $\bf{r_{v4}}$ and $\bf{r_{pfc}}$  minimizing the full cost function $E$  (Eq.\ref{eq:cost}), which is equivalent to maximizing the full posterior distribution in Eq.\ref{eq:posterior} composed of both the feedforward, $\kappa$-dependent distribution and the feedback, prediction-driven distribution. In this way, the model draws a connection between the response dynamics of V4 and PFC neurons and different computational stages in the feedforward-feedback loop. 

The inferred optimal responses of each neuronal unit in V4 and PFC across a range of occlusion levels, before and after the feedback from PFC, are shown in Fig.~\ref{fig:Result_Main}C,D, and E. Both PFC unit 1 and unit 2 responses increase with added occlusion (Fig.~\ref{fig:Result_Main}C), in agreement with the experiments where PFC neurons respond strongly to occluded stimuli and weakly to unoccluded stimuli (Fig.~\ref{fig:Experiment}D). Such increased PFC responses to occlusion result from the PFC connections to the occluder-selective V4 unit 3;\Ecommentnew{can you remove part of this sentence from in addition to the end?  And then link this and the next sentence with a colon} through the synaptic connections, PFC predictions are compelled to match the responses of V4 unit 3 which responds preferentially to occluders. The model PFC units also show shape selectivity, with PFC unit 1 showing higher responses than PFC unit 2 to the test shape A across occlusion levels. This agrees with physiological evidence for shape selectivity in PFC \cite{Pasupathy15}.

The two-step inference on the V4 responses accurately predicts the response characteristics of the initial and the delayed peaks in experimental recordings of V4 neurons. While the responses of V4 unit 2 (the neuronal unit not preferring the test shape A) stay constant at a low rate across the occlusion levels, V4 unit 1 (the preferred V4 unit) shows a decreasing response pattern as occlusion increases, i.e., as unoccluded area decreases. Compared to the responses inferred only based on the feedforward sensory input (Fig.~\ref{fig:Result_Main}D, solid green), the firing rates are less dependent on occlusion level when the feedback predictions are included (Fig.~\ref{fig:Result_Main}D, dotted green). Thus, with the feedback, an increase in occlusion does not as extensively degrade the preferred V4 responses. The model predictions therefore agree with the experimental observation on the two transients in V4 (Fig.~\ref{fig:Experiment}B), and are in accordance with our hypothesis that  the initial V4 responses reflect the feedforward signals from the afferent areas, and the delayed peak of responses in V4 are computed based on both the feedforward sensory signals and the feedback predictions from PFC. Because the response of the preferred V4 unit becomes is resistant to occlusion when the feedback prediction is included, we say that the feedback enables V4 neurons to have enhanced shape discriminability under partial occlusion.

Finally, the group of neurons that are hypothesized to respond preferentially to occluder saliency exhibits increasing responses as occlusion increases, both with and without the feedback (Fig.~\ref{fig:Result_Main}E). Although this class of neurons has not been systematically recorded in experiments, neurons selective for specific colors of occluders are known present in V4 \cite{Zeki73,Bushnell12}. 

In the above we have compared the steady-state representation of  neuronal responses in the model to transient peaks of responses in the experiments. 
%The dynamics of the neuronal responses in Eq. \ref{eq:descent} \Ecommentnew{should be consistent on whether or not have space before eq number} predict that the responses converge to the optimal steady-state values. 
The two-step inference does not have a mechanism for the shape of the transient activities observed in experiments. Specifically, instead of having the brief suppression of responses between the initial peak and the delayed peak (Fig.~\ref{fig:Experiment}A), the gradient descent on $E_1$ (Eq.\ref{eq:cost1}) and $E_2$ (Eq.\ref{eq:cost2}) with respect to $\mathbf{r_{v4}}$ simply predicts the V4 response dynamics $\mathbf{r_{v4}}$ to reach and stay at the respective steady state firing rates which minimize $E_1$ and $E_2$. 
%Likewise, gradient descent on $E_2$ with respect to $\mathbf{r_{pfc}}$ does not explain the transient dynamics of PFC responses, as once $\mathbf{r_{pfc}}$ reaches the steady state values that minimize the cost function, the firing rates stay at these values rather than subsequently being suppressed as in the experimental recordings. 
This implies that there may be additional physiological mechanisms-- for example, rapid suppression-- in the cortical circuitry responsible for the transient dynamics.   We note that, in principle, it is also possible that such temporal effects could also be interpreted by extending the predictive coding to the temporal domain \cite{Rao99a,Friston09a,Friston09b}.  \Ecommentnew{nice if any refs can give on end there, not crucial.  Also I shortened above, please make sure still OK, several sentences seemed redundant}  %Implementation of temporal predictive coding, however, is beyond the scope of this study. 

In summary, in this section we asked how the responses in a hierarchical predictive coding model compare to physiology.  We find that, upon training, the model indeed predicts the observed responses in V4 and PFC, when the dynamics unfold over an initial feedforward and a second  feedback stage.\Hcomment{Removed}\Ecomment{do we want to make this robustness point here or save for later?} \Acomment {you and Eric can decide but to me it seems that the results so far read like a lot of methods with a sprinkling of results.\Hcomment{Yes it feels a little redundant, but we wanted to give a brief recap of the Methods for the readers who skip the method section and jump right onto Results} Why does this section just not start by saying our network architecture captures observed results. When we trained the network it converged to these weights and then when we derived responses in PFC and V4 they matched physiology? Its buried in there but maybe more powerful to state it up front.\Hcomment{I edited the first part of this section to address this.}}

\subsection*{Parsimony of the network structure}

In the simulations above, we have assumed a specific network structure. This poses the question of whether these assumptions were necessary, and in general what aspects of network structure are required to reproduce the observed physiological responses.

Shape selectivity in V4 and PFC neurons is supported by experiments, thus we included the test shape-preferred and non-preferred V4 and PFC units, namely, V4 units 1 and 2 and PFC units 1 and 2. In addition, our model includes an additional group of V4 cells that responds strongly to occlusion. \Hcomment{Revised}\Ecomment{wonder if can add some phrases to make this section read a bit more results-y.  e.g., we found that ... the reasoning below shows that ... we experimented with different architectures and found that ... whatever is on target and seems to help!  Applies to this and the next 2 paras}We found that such occluder-selective V4 neurons are necessary to capture the response characteristics of PFC neurons observed in the experiments. Since the second term in the cost function Eq. \ref{eq:cost} is the squared difference between the PFC predictions -- a linear combination of PFC responses -- and the actual V4 responses, the PFC responses minimizing the cost function tend to follow the response trends of the afferent V4 neurons. The shape A (test shape)-preferred V4 unit 1 exhibits monotonically decreasing firing rates as occlusion level increases, while the activity of the shape B-selective V4 unit 2 stays constant across degrees of occlusion, as a consequence of the bottom-up stimulus-dependent inputs. With only these two types of neuronal populations, therefore, the PFC responses cannot capture the firing rate increase induced by occlusion. Given our model architecture without any additional mechanisms, there has to be a class of V4 neurons that responds strongly to occlusion but only weakly to unoccluded stimuli, so that PFC follows the similar response trends.  Moreover,  we found that the increase in PFC responses with occlusion cannot be obtained by including by simple prior distribution in the cost function instead of the third class of V4 neurons in question \Hcomment{I guess the occlusion sensitive V4 neurons are necessary ``in our proposed architecture'', which does not include any additional mechanisms other than simple hierarchical inference. Added this condition above in a couple sentences. }\Acomment{I would soften this. Are the occlusion sensitive V4 neurons necessary? FOr example, could we get this by normalization ? Or V4 neurons inhibiting PFC neurons (via inhibitory inter-neurons?)}

Another feature of our architecture -- the convergence of the signals, with each of the PFC cells connected to multiple afferent V4 neurons from different populations -- is also critical to replicate the \Hcomment{Changed}\Ecomment{will readers understand what improved means here?  else rephrase ...} shape selective responses that become more robust to occlusion after the PFC feedback. We experimented with different architectures and found that such convergence is crucial for transmitting information between different V4 units. Unless the same PFC unit makes predictions about both the shape-preferred V4 unit (V4 unit 1) and the occluder-selective V4 unit (V4 unit 3), the information about the occlusion level encoded by the occluder-selective V4 unit will not be transmitted to the shape-selective V4 population, which is crucial for maintaining robust shape discrimination and weaker dependence on occlusion. This structure, where the neurons of the lower cortical areas with different tuning properties send convergent signals to neurons in higher cortices, agrees physiological findings in which signals become more mixed as they travel along the hierarchy \cite{Felleman91,Rigotti13,Fusi16}.
%In addition, \Ecomment{active phrasing ... we find that ... or something} the model setup requires fewer active units in the higher area compared to the lower area. 

Another feature of our model is that fewer units in PFC (2) combine to make linear predictions about the responses of a larger number (3) of  V4 units.  This is also necessary to capture the experimental data.  Without such  convergence, the V4 responses imposed by the bottom-up sensory input can be matched perfectly by the top-down predictions made by PFC units, leading the optimal predictive coding solution to make identical copies of the sensory input at each stage along the hierarchy -- which clearly does not occur in experiments.  Translating this constraint into biology, this does not mean there must be fewer neurons in higher areas of brain, but rather that there are fewer functional or active populations that can be grouped as single units in the higher area during the task.  

In summary, the proposed network, composed of two PFC units and three V4 units, has a parsimonious structure to explain the neuronal responses in the experiments under  predictive coding principles.

\subsection*{Differential weighting of feedforward and feedback inputs}

In our model, the relative strength of feedback and feedfoward interactions are determined by assumptions about levels of variability in the inference errors (the noise terms in Eq.\ref{eq:noise1},\ref{eq:noise2}) at each network layers (Eq.\ref{eq:variance1},\ref{eq:variance2}). \Ecommentnew{some rephrasing, please check}Here we ask how these assumptions impact the ability of the model to reproduce  trends in experimental data. \Hcomment{Rephrased}\Ecomment{maybe last word should be more specific, trends in experimental data or recordings or something} 
 
Recall that the cost function $E$ in our model has two terms, one based on bottom-up sensory inputs and the other based on top-down predictions (Eq.\ref{eq:cost}). Contribution of each of these components is weighted by the inverse variance of the respective probability distribution. The pattern of the optimal responses to occlusion can therefore be modulated by these variances. Here we examine how this occurs, 
%patterns of the model units as we change the amount of relative contributions of the feedforward and feedback components by modulating the variances, 
and show that the tradeoff between feedforward and feedback components achieved by the variances in Fig.~\ref{fig:Result_Main} is necessary to capture the response characteristics observed in experiments.

\begin{figure}[!htb]%[h]
\noindent \centering
\includegraphics[scale=0.7]{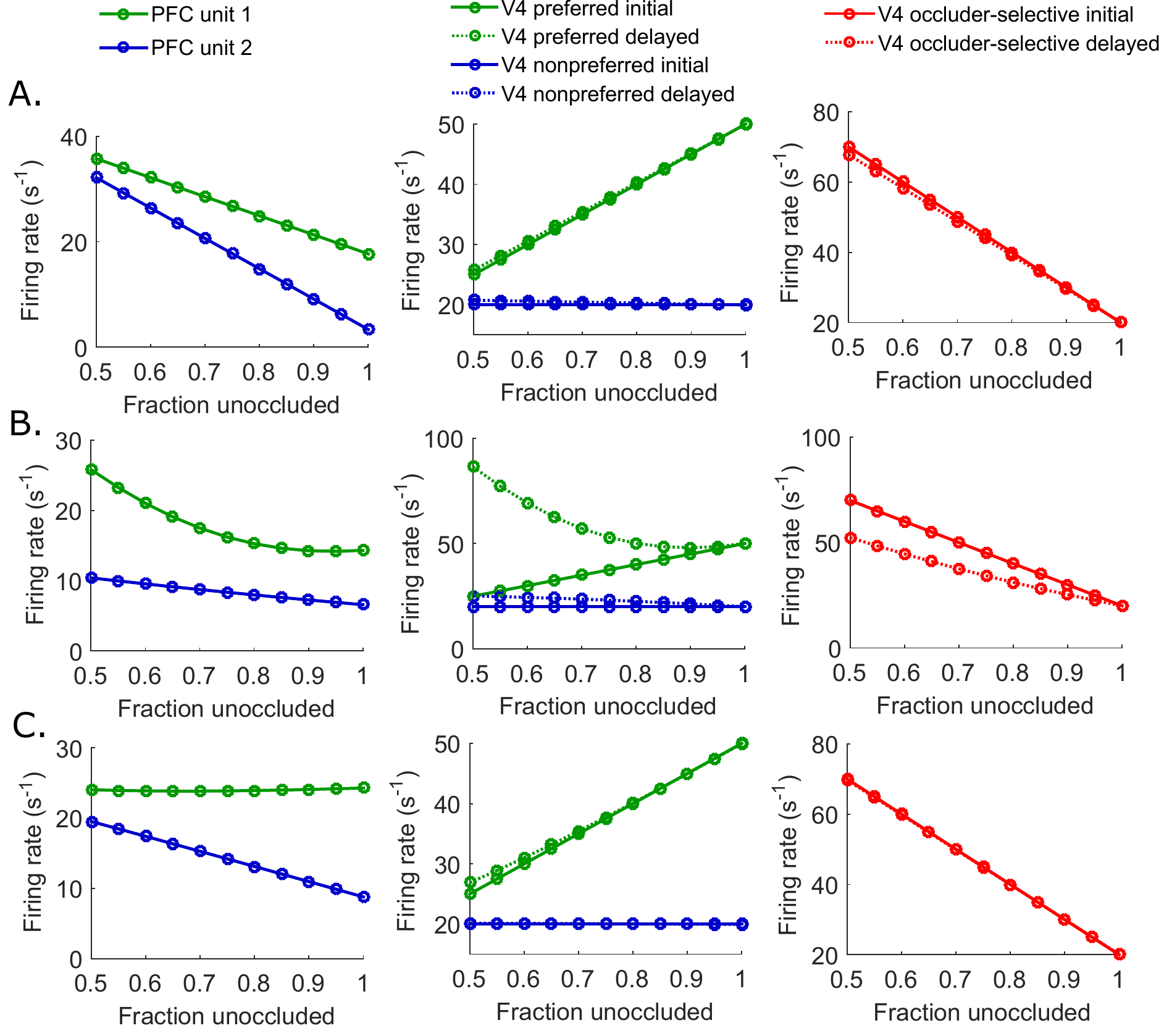}
\caption{{\bf Model simulations with modified top-down and bottom-up variances predict different response patterns in neuronal units} 
The responses of each neuronal unit when (A) the bottom-up variance of shape A-selective V4 response distribution $\sigma_1$ stays constant with increasing occlusion, (B) the top-down predictive distributions all have unit variances  ($\sigma'_1=\sigma'_2=\sigma'_3= 1$), (C) the top-down variances are all larger than the bottom-up variances ($\sigma'_1=\sigma'_2=\sigma'_3= 10$). }
\label{fig:Variance}
\end{figure}

We first discuss effects of the variances for the bottom-up input-driven distributions. In the original model (Fig.~\ref{fig:Result_Main}), for the bottom-up component,  variances are set equal to 1 for all three V4 populations when the input shape is unoccluded. We also set the variance for the test shape-preferred V4 population (V4 unit 1) to increase as occlusion level increases, to capture the increase in uncertainty of the shape identity in presence of occlusion. We found \Hcomment{Rephrased}\Ecomment{here more active phrasing used, important to try to change this throughout ... } that this increase in variance for the preferred V4 unit is necessary to mimic its weaker sensitivity to occlusion when feedback inputs are included \Ecomment{lots of words cut here ... please try to do this type of cutting in other sentences, which unless I'm screwing things up will make them easier to read.}. Without the increase in variance, this V4 unit depends relatively more on the bottom-up inputs under high degrees of occlusion \Hcomment{If we don't have beta=5 for V4 unit 1 so that the variance for V4 unit 1 stays the same, the ``weight'' on the bottom-up term for this unit does not get smaller under high levels of occlusion, so even under high occlusion, the response gets controlled mostly by the bottom-up term. Rephrased to clarify. }\Ecomment{why does this effect occur specifically at **high** levels of occlusion?  if I'm reading Beta right on page 9, isn't var of occluder pop constant, so that balance between terms would seem to be about the same in this case across occlusion levels?  I'm likely missing something ...}, and as a result, shows a steep decrease in its responses as occlusion increases (Fig.~\ref{fig:Variance}A, middle panel, green). By increasing the variance of the sensory input-dependent distribution, therefore, the optimal response of this V4 population becomes more dependent on the top-down predictions made by PFC. As the PFC populations respond strongly to occluded stimuli, weighting the bottom-up component less will result in a more gradual decrease in V4 responses to increasing occlusion, as in the original model in  Fig.~\ref{fig:Result_Main}D.

%%%%%%%%%%%%%%%%%%%%%%%%%%%%%%%%%%%%%%%%%%%%%%%%%%%
%%%%%%%%%%%%%%%%%%%%%%%%%%%%%%%%%%%%%%%%%%%%%%%%%%%

Next, we examine the choice of the top-down variances in the original model that successfully captures experimental data. In the initial model (Fig.~\ref{fig:Result_Main}), the variances of the top-down component do not depend on the occlusion level and stay at constant values.  However, the top-down effect is differentially weighted for each of the V4 populations; it is weighted more for the occluder-selective V4 population ($\sigma'_3= 1$) compared to the shape A- and B-selective neurons ($\sigma'_1= \sigma'_2 = 10$). This is needed to reproduce the rise in PFC responses at higher levels of occlusion. %The occluder-selective V4 responses increase with occlusion. 
The smaller variance, or equivalently, more ``weight", on the top-down predictions of the occluder-selective V4 unit drives the PFC unit to follow the same increasing response pattern as the occluder-selective V4. The smaller variance imposed on the top-down prediction \Hcomment{Rephrased}\Ecomment{grammar / phrasing issue here .. variance of effect (?)} for the occluder-selective V4 unit can be interpreted as the top-down predictions having more significance for occlusion than for identity of the shape. \Hcomment{Right, it is an assumption so should be rephrased. I think we can just remove these sentences.  }\Ecomment{previous two sentences ... during training, U changes.  Is that what is responsible for the effect just mentioned?  Or is it the ad hoc assumption about about the sigma term?  Should clarify ... and probably if it is just a choice we made say that more directly without mentioning training, which could be confused with things that actually do update during training}

\Hcomment{Revised, previous paragraph on the original values, and this paragraph on changing the variances}\Ecomment{This reads as the same topic sentence as the previous para.  I am probably missing something.  Otherwise need to do some writing to make flow of ideas / work more clear.} 
We investigated effects of changes in the top-down variances on the response patterns. When the feedback prediction-driven distributions for all V4 units are uniformly weighted with unit variance, the top-down effect becomes more pronounced (Fig.~\ref{fig:Variance}B) compared to the case with the variances at the original values (Fig.~\ref{fig:Result_Main}D).  As a consequence, the delayed responses of the test shape-preferred V4 (V4 unit 1) increase with added occlusion, reflecting strong modulation by PFC (Fig.~\ref{fig:Variance}B, middle panel, green dotted line). Similarly, when the top-down variances on all three V4 units are set to be larger than the bottom-up variances, relatively more influence is exerted by the bottom-up drive (Fig.~\ref{fig:Variance}C). As a result, the feedback no longer increases robustness of V4 unit 1 responses under partial occlusion (Fig.~\ref{fig:Variance}C, middle panel, green dotted line).

In sum, we have shown that the ability to reproduce trends in experimental recordings in our predictive coding model requires the balance of top-down and the bottom-up influences that is given by the increase in the input-dependent variance with added occlusion for the test shape-selective neurons and the smaller variance in the top-down prediction on the occluder-selective neurons. \Ecommentnew{fill in phrase specifically describing asymmetry} \Hcomment{This paragraph reads to me as a concluding/summarizing remark for this section, as it talks only about the feedback and feedforward effects balanced by modulating the inverse variances.}\Ecomment{Do you think this whole para could be moved to discussion?}  \Ecommentnew{not end of world but I still think this would be better off in discussion ... future experiment is not a result really}

\subsection*{Model prediction for responses to non-salient occlusion, noise, or reduced contrast}

\begin{figure}[!htb]%[h]
\noindent \centering
\includegraphics[scale=0.65]{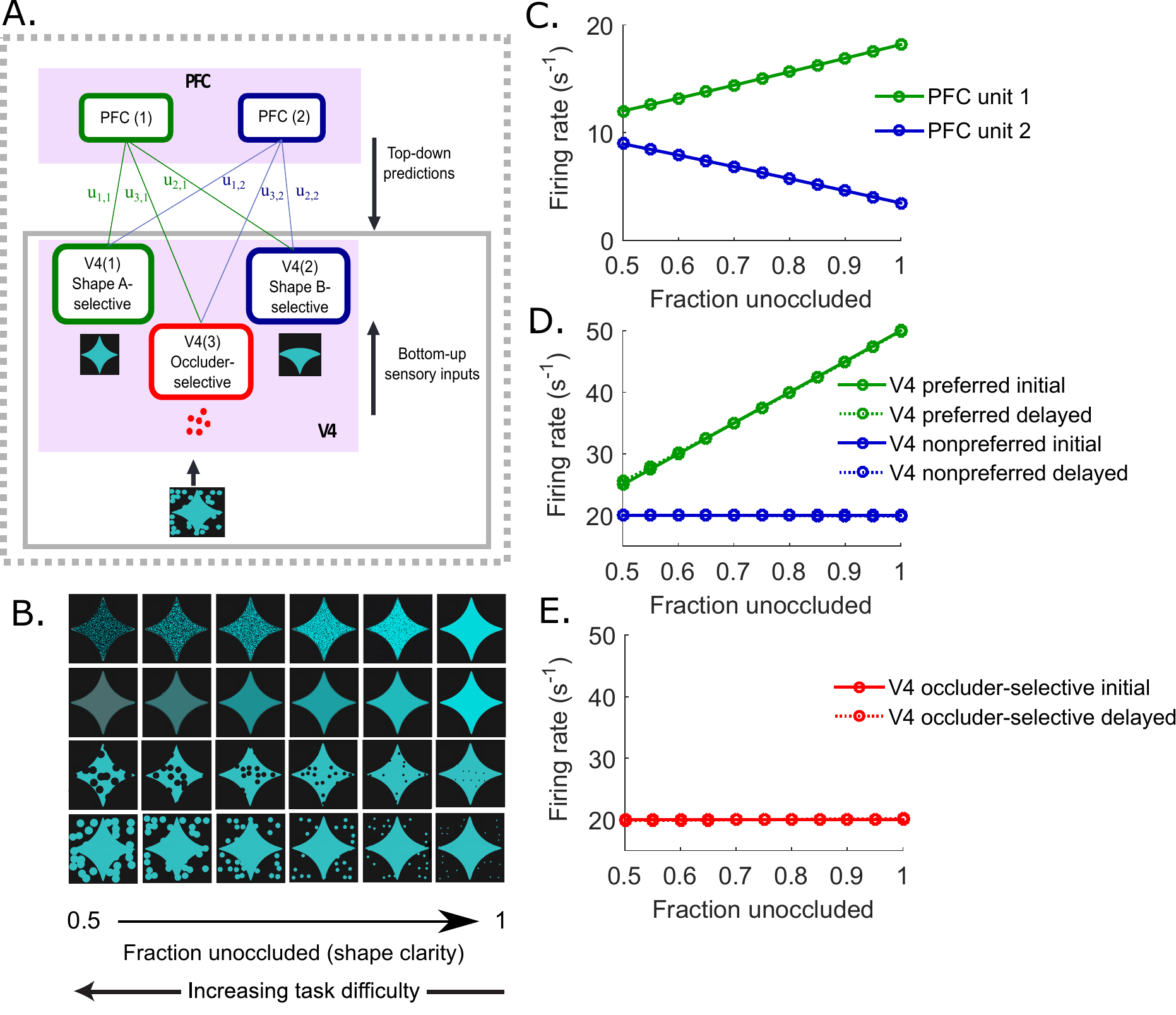}
\caption{{\bf Model simulation with indiscriminate occlusion or noise does not activate a class of V4 neurons, predicting the top-down signals to have no effect on the V4 responses.}
(A) Model schematic. Same model as in Fig.~\ref{fig:Result_Main}A, but with an input stimulus obscured by non-salient occlusion, noise, or reduced contrast.  \Hcomment{Rephrased}\Ecomment{awkward phrasing at end of sentence, though I couldn't manage to improve myself!} (B) Illustration of the input stimuli:  shape A with varying degrees of noise, contrast, and non-salient occlusion with occluders of the same color as the background or the shape. These types of visual ambiguity are not salient while obscuring the shape identity.  (C) Inferred PFC responses as a function of fraction of the shape unoccluded (shape clarity  \Ecomment{aha!  clarity is a nice word, maybe can use to smooth or compress some of the phrasing in the text also}). Reduced shape clarity alone \Hcomment{Rephrased}\Ecomment{would read easier if do not list all 3 possibilities here} does not increase the responses of shape A-selective PFC population. (D) Inferred responses of the shape-selective V4 units before (solid) and after (dotted) the top-down prediction, as a function of occlusion/obscurity level. The responses are depicted by color and line type as in Fig.~\ref{fig:Result_Main}D. The responses of the preferred V4 population after the top-down inputs are not distinguishable from those before the top-down inputs. Therefore, the top-down prediction does not improve shape discriminability under occlusion. (E) Model prediction of average firing rates of the occluder-selective V4 population. The non-salient occlusion does not activate the V4 population selective for some distinct feature (e.g. color) of the occluders. Note that fraction unoccluded on the x-axis means shape clarity in the case of reduced contrast or added noise.}
\label{fig:Result_Nonsalient}
\end{figure}

Above, we have assumed that occlusion is salient, and that there is a separate population of cells in V4 that responds preferentially to occlusion.  But what happens to predictions of the model when the occlusion is non-salient -- that is, indistinct from the shape? To answer this, we consider the case where the occluder reduces the shape signal, but does not activate a dedicated class of V4 neurons. For example, when the occluders are of the same color as the shape or the background, occlusion would increase ambiguity of the shape identity but would not induce responses in a V4 population separately responsive to a distinct color. Other examples include a decrease in shape clarity by white noise or reduced contrast (illustrated in Fig.~\ref{fig:Result_Nonsalient}B). 

%%%%%%%%%%%%%%%%%%%%%%%%%%%%%%%%%%%%%%%%%%%%%%%%%%%
%%%%%%%%%%%%%%%%%%%%%%%%%%%%%%%%%%%%%%%%%%%%%%%%%%%

We simulated such non-salient occlusion and ambiguity in our model by setting $\mu_3$, and therefore the peak of the response distribution for the occluder-selective V4 conditioned on sensory stimulus, to a constant. Therefore, an increase in occlusion or ambiguity in the shape stimulus does not increase the responses of V4 unit 3, as shown in Fig.~\ref{fig:Result_Nonsalient}E.  The peak $\mu_1$ for the shape A-preferring V4 unit, however, is assumed to decrease with occlusion, as for previous simulations.  This results in a decrease in the preferred PFC responses with occlusion/ambiguity, and only a slight increase in the non-preferred PFC responses (Fig.~\ref{fig:Result_Nonsalient}C).\Hcomment{Is the revised sentence here better?}\Ecomment{phrasing in prev few words confusing ... maintain, robust} Therefore, the feedback predictions made by PFC do not increase the preferred V4 unit 1 responses when the shape ambiguity (occlusion level) is high. In Fig.~\ref{fig:Result_Nonsalient}D, the preferred V4 responses after the feedback (dotted green) are therefore indistinguishable from the responses before the feedback (solid green). Our model thus predicts that when the shape signal is occluded in a way that is not salient, the feedback from PFC does not improve shape discriminability. 

From the point of view of perception, this prediction seems plausible since we often have more difficulty recognizing an object when the obscurant is not distinct from the object. Moreover, preliminary experimental observations show that PFC neurons do not respond strongly to occluders of the same color as the background. In addition, the second peak of responses were not observed in V4 neurons when the shapes were obscured by reducing their contrast. While these preliminary observations are in accordance with our model predictions, more data should certainly be collected before conclusions can be drawn.

\section*{Discussion}

In this study, we have proposed that robust shape-selective V4 responses under partial occlusion can be explained in the framework of predictive coding and hierarchical Bayesian inference.  We have used this framework to construct a model of V4 and PFC in which signals converge as they travel up the hierarchy. In particular, we suggest that top-down predictions made by PFC neurons with mixed selectivity for shape identity and occlusion play a significant role in maintaining robust shape discriminability under partial occlusion in V4.  In this model, PFC neurons make linear predictions on V4 activities in the form of feedback signals, and the connection weights are interpreted to store the memory of the shape identities. We reformulated the traditional framework of predictive coding, so that the optimal representation of the internal states of the model V4 and PFC units, rather than residual errors, are comparable to the electrophysiological recordings in these areas. 

Our model suggests that the initial responses in experimental recordings of V4 are purely feedforward and computed solely based on the bottom-up sensory input, while the delayed responses are modulated by both the bottom-up sensory signals and the top-down predictions. The model further shows that the feedback signals in V4 improve the shape discriminability under occlusion by reducing ambiguity in the population representation of the shape identity, and that this is achieved by transmission of the occlusion information via a feedforward-feedback loop. This can be viewed as an extension of the concept proposed in \cite{Rao99a} where predictions made by higher visual areas with larger receptive fields enable neurons encoding the surround and the center in V1 to share information; in our model of V4, neurons encoding different features of a shape stimulus such as curvature, color, etc, share information via predictions made by the higher areas. 

The increase in the shape selective responses of V4 induced by the feedback depends on asymmetric weighting of the top-down and the bottom-up effects, so that the top-down prediction is weighted more strongly for the occluder-selective neurons and the dependency of the shape-selective neuronal responses on the sensory input decreases with added occlusion.\Ecommentnew{I'd be more specific here with a few more words --- something like ... with variances chosen so that xx vs xx weighted more strongly. } Interesting future work could more directly  test this weighting of the top-down and the bottom-up effects. For example, weakening the top-down predictive component by either training with larger set of noisy shape stimuli, or possibly by cooling PFC, might result in more emphasis on the bottom-up sensory input and thus a smaller increase in in shape selectivity during delayed V4 responses. 

\Ecommentnew{Hannah ... how about working this in as ... a concluding sentence with some modificaitons maybe in its own para} In this way, our model contributes to new understanding of both neurophysiological and computational mechanisms underlying discrimination of partially occluded shapes in V4, suggesting a possible functional contribution of feedback signals.
%Our model, therefore, gives insight for implementing experiments needed for better understanding of the information processing in the intermediate and higher visual areas, in addition to suggesting a computational principle underlying putative feedback signals to V4.   

\subsection*{Relationship to previous models}

Several previous theoretical studies investigated the computational mechanisms for recognition of partially occluded shapes, patterns, and objects \cite{Fukushima87, Fukushima01, Fukushima05, Rao97}.  However,  these are strictly feedforward and often overlook feedback computation, in stark contrast to biological networks which feature abundant feedback and recurrent connections.
One approach is based on an extended version of neocognitron-- a hierarchical,  multilayered, and feedforward neural network model \cite{Fukushima87, Fukushima01, Fukushima05}. This extended neocognitron has an additional ``masker layer" which detects occluders by difference in brightness and suppresses them at an early state.  A study by Rao \cite{Rao97, Rao99b} uses a Kalman filter model and Bayesian optimal estimation theory of maximizing the posterior probability of the internal states. With robust optimization method which clips large residual errors, the model effectively segments the occluders from the image, treating the occluders as the outlier. The physiological mechanisms  \Hcomment{correspondence? mechanisms?}\Ecomment{rephrase ``relevance'' to be more specific} underlying the robust optimization method, however, are not known. 

There have been a number of other modeling studies of V4 tuning to shape contours based on hierarchical feedforward models of object categorization, which have structural similarity to the ventral visual pathway  \cite{Fukushima80, Riesenhuber99, Serre07, Cadieu07, Yamins14}. These models are also purely feedforward, and while they have had successes in reproducing V4 shape selectivity \cite{Cadieu07, Yamins14}, they lack separate mechanisms to account for occlusion.  Unlike these previous models, our model bridges hierarchical predictive coding,  and experimentally recorded response dynamics in area V4 and PFC.

\subsection*{Information transmission through feedforward-feedback and recurrent connections}

In our model, all the information about the input stimulus, namely, the shape identity $s$ and the occlusion level $c$, is already available at the V4 level. This information is represented by the response distributions of multiple V4 populations given the input $\kappa = (s,c)$.  This suggests a natural question. If all the necessary information is already present in V4, why does the system implement the feedforward-feedback loop and involve the higher area PFC for shape discrimination?

\begin{figure}[!htb]
\noindent \centering
\includegraphics[scale=0.8]{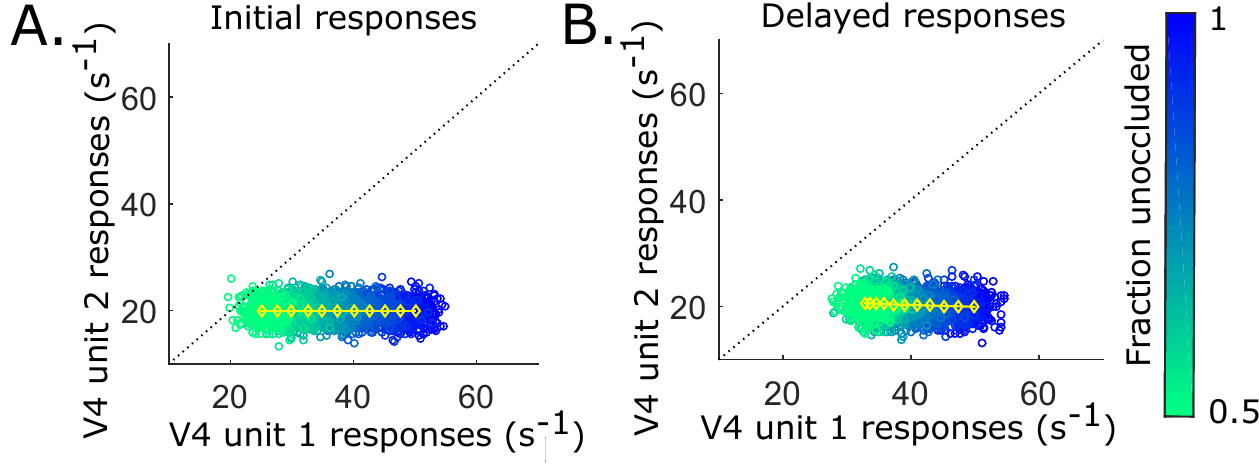}
\caption{{\bf Shape discriminability under occlusion increases with the top-down prediction}  
The optimal average firing rates across degrees of occlusion as in Fig.~\ref{fig:Result_Main}D (yellow),  projected onto the state space of V4 unit 1 (preferred) and unit 2 (non-preferred) responses. For each occlusion level, 200 responses were generated with a white noise with the mean at the optimal average value (yellow) and standard deviation of 2 arbitrary chosen for illustration purpose  (blue: low occlusion, green: high occlusion) \Hcomment{Addressed}\Ecomment{ Let's try to write this really directly to make sure nobody gets confused about origin of noise here ...  e.g., explaining that the noise std dev of 2 is arbitrary for illustration purposes and has nothing to do with sigmas}. When the population responses are under the unity line (dotted black), $r_{v4,1}>r_{v4,2}$, and the animal concludes that the test shape presented is shape A. The opposite is true for $r_{v4,2}>r_{v4,1}$. Before the top-down prediction (A), the noisy responses under high occlusion (green dots) lie close to the unity line, obscuring the shape identity. With the top-down prediction included (B), the average optimal responses to occluded stimuli are moved horizontally to larger $r_{v4,1}$ values (yellow). Thus the noisy responses are more squeezed and moved away from the unity line, clarifying the shape identity. 
}
\label{fig:Result_StateSpaceView}
\end{figure}

\Hcomment{Done}\Ecomment{need some sort of a transition sent that indicates are answering this main question next} To answer this question, we first examine how the visual input with partial occlusion is represented in V4 neurons. The increased shape selectivity under occlusion during the delayed responses is illustrated in a state space view in Fig.~\ref{fig:Result_StateSpaceView}. Based on previous experimental evidence  \cite{Kosai14}, we assume that shape recognition is performed by comparing the population responses of shape A-selective and shape B-selective V4 neuronal groups ($ r_{v4,1}>r_{v4,2}\rightarrow \text{shape A}$, $ r_{v4,1}<r_{v4,2}\rightarrow \text{shape B}$).
\Hcomment{Revised}\Ecomment{rewrite with a topic sentence for para ... flow in this para and link with previous one could be imrpoved}Our simulations with partially occluded shape stimuli found the optimal firing rates of each neuronal population (Fig.~\ref{fig:Result_Main}D). These firing rates are projected onto the state space of V4 unit 1 and unit 2 responses in Fig.~\ref{fig:Result_StateSpaceView} (yellow). For each occlusion level, 200 responses were generated with a Gaussian noise around the optimal firing rates. As occlusion increases, the responses move towards the unity line (black dotted line). Without the feedback predictions included in Bayesian inference, during the initial responses, high occlusion moves noisy versions of the responses close to, or even above, the unity line, obscuring the shape identity (Fig.~\ref{fig:Result_StateSpaceView}A). However, when the feedback from PFC is included, the responses move away from the unity line, thus clarifying the shape identity under partial occlusion  (Fig.~\ref{fig:Result_StateSpaceView}B). 

%%%%%%%%%%%%%%%%%%%%%%%%%%%%%%%%%%%%%%%%%%%%%%%%%%%
%%%%%%%%%%%%%%%%%%%%%%%%%%%%%%%%%%%%%%%%%%%%%%%%%%%
The convergent structure of the network is the key for this effect to occur. Although the information of occlusion level is present at the level of V4, it does not impact the shape selective V4 units without the feedback from PFC. In other words, the PFC predictions re-map the information about the shape identity and the occlusion level onto the shape-selective V4 space, enhancing the shape discriminability in V4. 

We note that recurrent connections among V4 populations -- rather than the feedback described above -- could in principle also transmit information about the occlusion level to the shape-selective neurons. \Hcomment{Deleted}\Ecomment{delete or compress prev sentence?} Which mechanism is more effective and efficient is an open question. However, the current experimental evidence showing the delayed peak of responses in V4 arising after PFC responses peak, as well as the strong PFC responses to occlusion, are both suggestive of the feedback mechanism. \Ecomment{compressed, could still use some smoothing over}

\subsection*{Learning the shape templates with connection weights}

Our model modifies the synaptic weights between V4 and PFC neurons during the preliminary phase which consists of a few presentations of unoccluded shapes. \Hcomment{This paragraph revised}\Ecomment{here and throughout please think about what details can be omitted ... e.g. we don't need to respecify 30 here, etc, but there are probably many other things that can be cut ... including the next sent too or most of it?} This step corresponds to the initial learning phase in experiments where the animal discriminates an unoccluded pair of shapes used for the session. In this setup, the fast learning of the shape pair after exposure to the shapes for just a few times, is achieved by the memory stored in the synaptic weights between V4 and PFC neurons. When partially occluded shapes are used during the preliminary phase, on the other hand, the system learns different values of synaptic weights and the feedback does not improve shape discriminability (See S2 Text, S2 Fig). Fast learning, as attested by the shape discrimination task here, has been observed widely, where new sensory stimuli are easily learned with just a few presentations \cite{Seitz10, Rubin97}.

Physiological recordings in cortical cells in vitro, however, show only small changes in synaptic strength after a pair of pre- and post-synaptic spikes \cite{Markram97, Bi98, Gerstner96}, suggesting that neurons learn a repeated stimulus more gradually, after a large number of presentations. Such seemingly contradicting evidence from physiology and behavioral observations can be reconciled by introducing stronger synaptic changes than usually observed in vitro, possibly aided by neuromodulation \cite{Fusi05}. More recently, it has been proposed that even weak synaptic plasticity can support fast learning in the balanced-regime of excitation and inhibition \cite{Yger15}. Due to the leverage effect from the excitatory and inhibitory balance in this regime, small synaptic modifications applied to many synapses onto a given neuron result in a large effect \cite{Yger15}.

\subsection*{Mapping computational nodes in predictive coding to cortical circuitry}

Different algorithms implementing hierarchical predictive coding share the general principle of a generative model:  the brain has an internal representation of the world which is actively compared to the actual sensory inputs. However, the precise computational procedures employed by these algorithms as well as their connections to neuronal populations are controversial and vary widely across different studies \cite{Spratling16, Bastos12, Bogacz15, Rao99a, Mumford92, Spratling08}. 

For example, in our model, the variances of the response distributions of different V4 units given the sensory input or the higher cortical activity are pre-defined to capture the response characteristics in experiments. However, they can also be treated as \Hcomment{Rephrased}\Ecomment{optimizing -- rephrase around here} parameters to be optimized and are assigned to the most likely values, with a slight modification on the network structure as done in a few other models of hierarchical predictive coding. In these studies, the variances are interpreted as synaptic weights and are obtained by minimizing the free energy \cite{Bogacz15,Friston09a,Friston09b}. 

There are varied interpretations on the connections between predictive coding algorithms and computations done by cortical circuitry. Cortical areas have laminar structures, and different layers or populations within the cortical area may correspond to different local computational nodes that arise in predictive coding algorithms. However, there is no unifying description of the intra-cortical connectivity and the local computations within a cortical area. For example, inhibitory feedback connection implemented in the model proposed by Rao et al \cite{Rao99a}, is modified in Spratling et al. \cite{Spratling08, Spratling16} to reflect excitatory feedback signals observed in physiology. In order to avoid negative responses, Spratling \cite{Spratling08, Spratling16} also replaced additive excitation and subtractive inhibition in \cite{Rao99a} by multiplicative and divisive modulations, respectively. In our model, we follow the approach in \cite{Rao99a} and implement additive excitation and subtractive inhibition for simplicity.

Within area V4, there surely are multiple neuronal populations across the laminar structures, and each neuronal node may perform different computations as suggested by earlier studies.  Investigations of specific neuronal populations within V4-PFC circuitry in the context of the corresponding computational nodes in the predictive coding algorithm will provide a better understanding and validation of our model.

\section*{Supporting Information}

\paragraph*{S1 Appendix.}
\label{S1_Appendix}
{\bf Population average responses.} This section provides a justification of modeling each population of V4 and PFC neurons as a single unit, based on a simulation with a group of slightly heterogeneous neurons for each neuronal unit in V4 and PFC .

\paragraph*{S2 Appendix.}
\label{S2_Appendix}
{\bf Connection weights learned with partially occluded shapes.} This section shows simulations with the connection weights tuned by training on shapes under partial occlusion, concluding that preliminary learning of unoccluded shapes is necessary for the feedback-induced enhancement in shape discriminability under partial occlusion. 

% Include only the SI item label in the subsection heading. Use the \nameref{label} command to cite SI items in the text.

%\subsection*{S1 Fig}
%\label{S1_Fig}

\section*{Acknowledgments}
We thank Rajesh Rao, Wyeth Bair, and Joel Zylberberg for many helpful discussions and comments. This work was supported by the Washington Research Foundation Innovation Postdoctoral Fellowship in Neuroengineering to H.C., NEI grant R01EY018839 to A.P., NSF Career Award DMS-1056125 to E.S-B, Vision Core grant P30EY01730 to the University of Washington, and P51 grant OD010425 to the Washington National Primate Research Center.\\

\clearpage

\nolinenumbers

%\section*{References}
% Either type in your references using
% \begin{thebibliography}{}
% Text
% \end{thebibliography}
%
% OR
%
% Compile your BiBTeX database using our plos2015.bst
% style file and paste the contents of your .bbl file
% here.
% 

% \begin{thebibliography}{10}
% \bibitem{bib1}
%Devaraju P, Gulati R, Antony PT, Mithun CB, Negi VS. Susceptibility to SLE in South Indian Tamils may be influenced by genetic selection pressure on TLR2 and TLR9 genes. Mol Immunol. 2014 Nov 22. pii: S0161-5890(14)00313-7. doi: 10.1016/j.molimm.2014.11.005
% \bibitem{bib2}
%\end{thebibliography}

\bibliography{HannahChoi}

\begin{thebibliography}{10}

\bibitem{Rust10}
Rust NC, Stocker AA.
\newblock {Ambiguity and invariance: two fundamental challenges for visual
  processing}.
\newblock Current Opinion in Neurobiology. 2010;20:382--388.

\bibitem{Gregoriou14}
Gregoriou GG, Rossi AF, Ungerleider LG, Desimone R.
\newblock {Lesions of prefrontal cortex reduce attentional modulation of
  neuronal responses and synchrony in V4.}
\newblock Nature Neuroscience. 2014;17:1003–1011.

\bibitem{Fukushima80}
Fukushima K.
\newblock {Neocognitron: A self-organizing neural network model for a mechanism
  of pattern recognition unaffected by shift in position}.
\newblock Biological Cybernetics. 1980;36:193–202.

\bibitem{Riesenhuber99}
Riesenhuber M, Poggio T.
\newblock {Hierarchical models of object recognition in cortex}.
\newblock Nature Neuroscience. 1999;2:1019 -- 1025.

\bibitem{Serre07}
Serre T, Oliva A, Poggio T.
\newblock {A feedforward architecture accounts for rapid categorization}.
\newblock Proc Natl Acad Sci. 2007;104:6424 -- 6429.

\bibitem{Cadieu07}
Cadieu C, Kouh M, Pasupathy A, Connor CE, Riesenhuber M, Poggio T.
\newblock {A model of V4 shape selectivity and invariance}.
\newblock J Neurophysiology. 2007;98:1733 -- 1750.

\bibitem{Yamins14}
Yamins DLK, Hong H, Cadieu CF, Solomon EA, Seibert D, DiCarlo JJ.
\newblock {Performance-optimized hierarchical models predict neural responses
  in higher visual cortex.}
\newblock PNAS. 2014;111:8619–8624.

\bibitem{Roe12}
Roe AW, Chelazzi L, Connor CE, Conway BR, Fujita I, Gallant JL, et~al.
\newblock {Toward a unified theory of visual area V4 }.
\newblock Neuron. 2012;74:12–--29.

\bibitem{Pasupathy99}
Pasupathy A, Connor CE.
\newblock {Responses to contour features in macaque area V4}.
\newblock J Neurophysiology. 1999;82:2490 -- 2502.

\bibitem{Pasupathy01}
Pasupathy A, Connor CE.
\newblock {Shape representation in area V4: position-specific tuning for
  boundary conformation}.
\newblock J Neurophysiology. 2001;86:2505 -- 2519.

\bibitem{Miller01}
Miller E, Cohen JD.
\newblock {An integrative theory of prefrontal cortex function}.
\newblock Annu Rev Neurosci. 2001;24:167–202.

\bibitem{Kosai14}
Kosai Y, El-Shamayleh Y, Fyall AM, Pasupathy A.
\newblock {The role of visual area V4 in the discrimination of partially
  occluded shapes}.
\newblock J Neuroscience. 2014;34(25):8570 -- 8584.

\bibitem{Pasupathy15}
Pasupathy A, Fyall AM, Choi H.
\newblock {Discriminating partially occluded shapes: insights from visual and
  frontal cortex.}
\newblock Cosyne annual meeting. 2015;.

\bibitem{Bogacz15}
Bogacz R.
\newblock {A tutorial on the free-energy framework for modelling perception and
  learning}.
\newblock Journal of Mathematical Psychology. 2015;.

\bibitem{Bastos12}
Bastos AM, Usrey WM, Adams RA, Mangun GR, Fries P, Friston K.
\newblock {Canonical microcircuits for predictive coding}.
\newblock Neuron. 2012;76:695–711.

\bibitem{Friston09a}
Friston K, Kiebel S.
\newblock {Predictive coding under the free-energy principle}.
\newblock Phil Trans R Soc B. 2009;364:1211–1221.

\bibitem{Friston09b}
Friston K, Kiebel S.
\newblock {Cortical circuits for perceptual inference}.
\newblock Neural Networks. 2009;364:1093–1104.

\bibitem{Srinivasan82}
Srinivasan MV, Laughlin SB, Dubs A.
\newblock {Predictive coding: A fresh view of inhibition in the retina.}
\newblock Proc R Soc Lond B Biol Sci. 1982;216:427—459.

\bibitem{Rao99a}
Rao RPN, Ballard DH.
\newblock {Predictive coding in the visual cortex: a functional interpretation
  of some extra-classical receptive-field effects}.
\newblock Nature Neuroscience. 1999;2:79--87.

\bibitem{Spratling16}
Spratling MW.
\newblock {A review of predictive coding algorithms}.
\newblock Brain and Cognition. 2016;.

\bibitem{Rao97}
Rao RPN.
\newblock {Correlates of attention in a model of dynamic visual recognition}.
\newblock Advances in Neural Information Processing Systems (NIPS). 1997;10:80
  -- 86.

\bibitem{Rao99b}
Rao RPN.
\newblock {An optimal estimation approach to visual perception and learning}.
\newblock Vision Research. 1999;39:1963 -- 1989.

\bibitem{Rao04}
Rao RPN.
\newblock {Bayesian computation in recurrent neural circuits}.
\newblock Neural Computation. 2004;16:1 -- 38.

\bibitem{Rao05}
Rao RPN.
\newblock {Bayesian inference and attentional modulation in the visual cortex}.
\newblock NeuroReport. 2005;16:1843 -- 1848.

\bibitem{Lee03}
Lee TS, Mumford D.
\newblock {Hierarchical Bayesian inference in the visual cortex}.
\newblock J Opt Soc Am A Opt Image Sci Vis. 2003;20:1434 -- 1448.

\bibitem{Yuille06}
Yuille A, Kersten D.
\newblock {Vision as Bayesian inference: analysis by synthesis?}
\newblock Trends in Cognitive Sciences. 2006;10:301--308.

\bibitem{Koch99}
Koch C, Poggio T.
\newblock {Predicting the visual world: silence is golden.}
\newblock Nature Neuroscience. 1999;2:9–10.

\bibitem{Meyers08}
Meyers EM, Freedman DJ, Kreiman G, Miller EK, Poggio T.
\newblock {Dynamic population coding of category information in inferior
  temporal and prefrontal cortex}.
\newblock J Neurophyiology. 2008;100:1407—1419.

\bibitem{Pasupathy02}
Pasupathy A, Connor CE.
\newblock {Population coding of shape in area V4.}
\newblock Nature Neuroscience. 2002;5:1332–1338.

\bibitem{Bushnell11}
Bushnell BN, Harding PJ, Kosai Y, Bair W, Pasupathy A.
\newblock {Equiluminance cells in visual cortical area V4}.
\newblock J Neuroscience. 2011;31(35):12398 -- 12412.

\bibitem{Rigotti13}
Rigotti M, Barak O, Warden M, Wang XJ, Daw ND, Miller RK, et~al.
\newblock {The importance of mixed selectivity in complex cognitive tasks}.
\newblock Nature. 2013;497:585--590.

\bibitem{Fusi16}
Fusi S, Miller RK, Rigotti M.
\newblock {Why neurons mix: high dimensionality for higher cognition}.
\newblock Current Opinion in Neurobiology. 2016;37:66--74.

\bibitem{Olshausen96}
Olshausen BA, Field DJ.
\newblock {Emergence of simple-cell receptive field properties by learning a
  sparse code for natural images}.
\newblock Nature. 1996;381:607–609.

\bibitem{Olshausen97}
Olshausen BA, Field DJ.
\newblock {Sparse coding with an overcomplete basis set: a strategy employsed
  by V1?}
\newblock Vision Research. 1997;37(23):3311–3325.

\bibitem{Zylberberg11}
Zylberberg J, Murphy JT, DeWeese MR.
\newblock {A sparse coding model with synaptically local plasticity and spiking
  neurons can account for the diverse shapes of V1 simple cell receptive
  fields.}
\newblock PLoS Comp Bio. 2011;7(10):e1002250.

\bibitem{Mumford92}
Mumford D.
\newblock {On the computational architecture of the neocortex}.
\newblock Biological Cybernetics. 1992;66:241 -- 251.

\bibitem{Zeki73}
Zeki SM.
\newblock {Colour coding in rhesus monkey prestriate cortex}.
\newblock Brain Res. 1973;53:422 -- 427.

\bibitem{Bushnell12}
Bushnell BN, Pasupathy A.
\newblock {Shape encoding consistency across colors in primate V4}.
\newblock J Neurophysiology. 2012;108:1299 -- 1308.

\bibitem{Felleman91}
Felleman DJ, Van~Essen DC.
\newblock {Distributed hierarchical processing in the primate cerebral cortex}.
\newblock Cerebral Cortex. 1991;1(1):1--47.

\bibitem{Fukushima87}
Fukushima K.
\newblock {Neural network model for selective attention in visual pattern
  recognition n and associative recall}.
\newblock Applied Optics. 1987;26:1985 -- 1992.

\bibitem{Fukushima01}
Fukushima K.
\newblock {Recognition of partly occluded patterns: a neural network model}.
\newblock Biological Cybernetics. 2001;84:251-- 259.

\bibitem{Fukushima05}
Fukushima K.
\newblock {Restoring partly occluded patterns: a neural network model}.
\newblock Neural Networks. 2005;18:33 -- 43.

\bibitem{Seitz10}
Seitz AR.
\newblock {Sensory learning: rapid extraction of meaning from noise.}
\newblock Curr Biol. 2010;20:R643–R644.

\bibitem{Rubin97}
Rubin N, Nakayama K, Shapley R.
\newblock {Abrupt learning and retinal size specificity in illusory-contour
  perception.}
\newblock Curr Biol. 1997;7:461–467.

\bibitem{Markram97}
Markram H, Lubke J, Frotscher M, Sakmann B.
\newblock {Regulation of synaptic efficacy by coincidence of postsynaptic APs
  and EPSPs.}
\newblock Science. 1997;275:213–215.

\bibitem{Bi98}
Bi GQ, Poo MM.
\newblock {Synaptic modifications in cultured hippocampal neurons: dependence
  on spike timing, synaptic strength, and postsynaptic cell type.}
\newblock J Neuroscience. 1998;18:10464–10472.

\bibitem{Gerstner96}
Gerstner W, Kempter R, van Hemmen JL, Wagner H, Hemmen JV.
\newblock {A neuronal learning rule for sub-millisecond temporal coding.}
\newblock Nature. 1996;383:76–81.

\bibitem{Fusi05}
Fusi S, Drew PJ, Abbot LF.
\newblock {Cascade models of synaptically stored memories.}
\newblock Neuron. 2005;45:599–611.

\bibitem{Yger15}
Yger P, Stimberg M, Brette R.
\newblock {Fast learning with weak synaptic plasticity.}
\newblock J Neuroscience. 2015;35(39):13351–13362.

\bibitem{Spratling08}
Spratling MW.
\newblock {Predictive coding as a model of biased competition in visual
  attention}.
\newblock Vision Research. 2008;48(12):1391–1408.

\end{thebibliography}

\clearpage

\section*{S1 Appendix. Population average responses}

\beginsupplement
In this appendix, the model is extended to include populations of neurons with slight heterogeneity. In our main model, each unit in the model V4 and PFC is considered as a population of neurons with similar tuning properties. For example, V4 unit 1 represents a population of V4 neurons that respond preferentially to shape A, and V4 unit 3 is interpreted as a population of V4 neurons responding strongly to some salient features of the occluders. With each unit representing a neuronal population, the optimal response inferred by minimizing the cost function depicts the average response of each neuronal population. Since the cost function in Eq.15 increases linearly with added neuronal units that share the same properties with the existing populations, representing a neuronal population as a single unit seems a reasonable simplification.

We now test this simplification explicitly.  We performed further numerical simulations with slightly heterogeneous group of neurons for each neuronal unit in V4 and PFC. The heterogeneity is introduced to the V4 neurons by assigning $\boldsymbol\mu$, the mean vector of the feedforward sensory input-driven response distribution, from a normal distribution with a unit standard deviation for each neuron within the population. Therefore, the bottom-up sensory input drives neurons within the same group to converge to slightly different optimal responses. In addition, the initial connection weights and the initial firing rates of the neurons are also slightly heterogeneous, chosen from normal distributions with the means at the initial values used in previous simulations, and the standard deviations of 0.1 for initial weights and 0.5 for initial firing rates. The PFC neurons in each population, therefore, also show weakly heterogeneous optimal representations as a result. Each neuronal population is composed of 10 slightly heterogeneous neurons. Moreover, each PFC neuron sends the prediction signals to one neuron from each of the three V4 populations, and each V4 neuron receives the feedback that is a weighted sum of two PFC neurons, one from each PFC population. Therefore, the convergence ratio from V4 to PFC is preserved as in the previous simulations where populations are represented as single units. We have tested a couple other convergence ratio (eg, two neurons per each V4 population connected to a single PFC neuron) and found that they produce the qualitatively similar results.

\begin{figure}[!htb]%[h]
\noindent \centering
\includegraphics[scale=0.63]{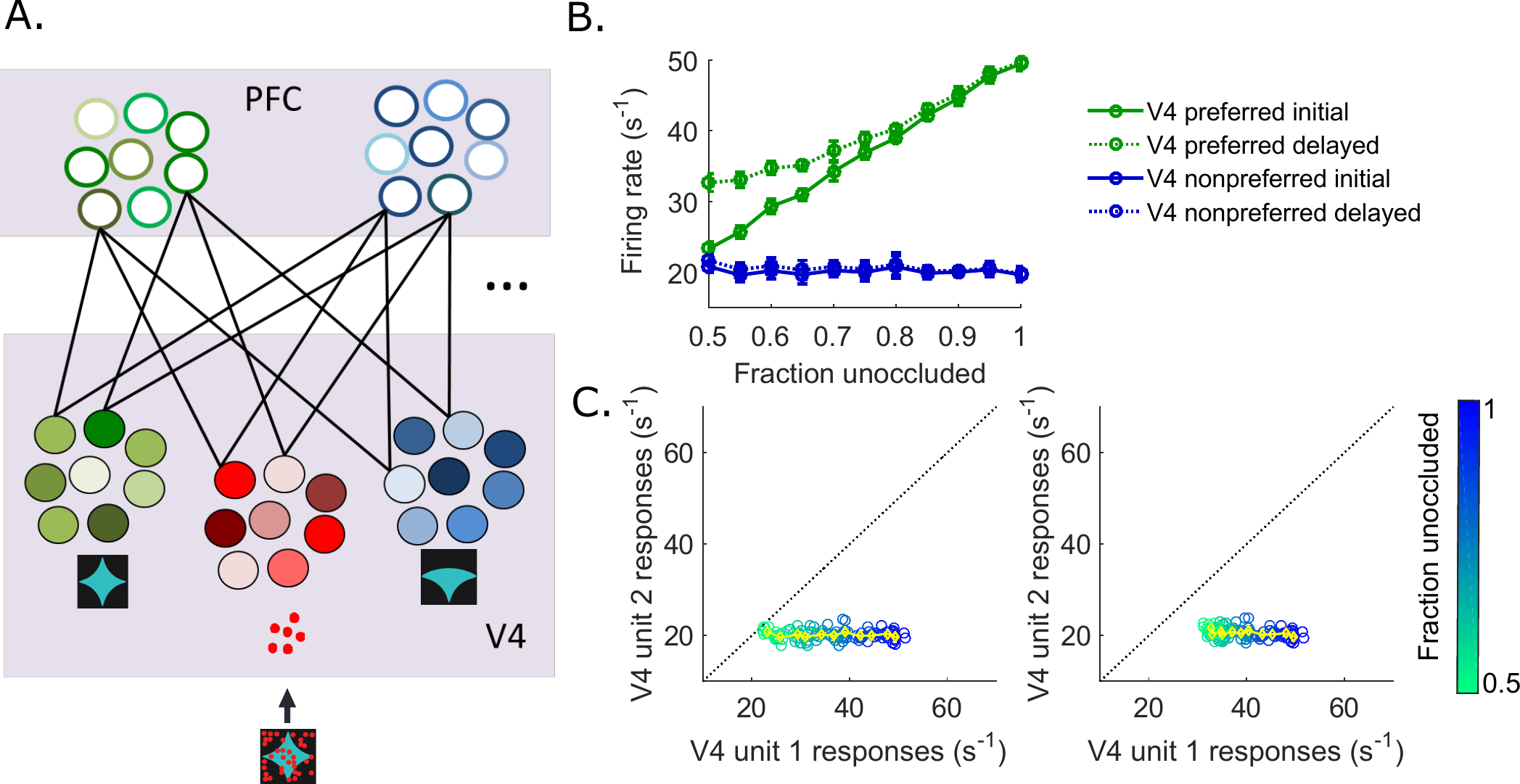}
\caption{{\bf  Simulation with slightly heterogeneous neurons within each population.} (A) Model schematic. For visualization, a smaller number of neurons per population and a subset of connections are shown. The actual model includes 10 neurons with similar tuning properties per population. Each neuron in PFC is connected to three V4 neurons from each V4 population, and each V4 neuron is connected to two PFC from each of the two PFC populations. The neurons of green shades prefer shape A and correspond to V4 unit 1, those of blue shades prefer shape B (V4 unit 2), and the neurons of red shades are selective for occluder properties (V4 unit 3). Varied shades of colors for neurons within each population represent slight heterogeneity. (B) Inferred responses of the shape-selective V4 neurons before (solid) and after (dotted) the top-down prediction. The green lines represent the optimal responses of the V4 population selective for the test shape A  and the blue lines are those of the non-preferred V4 population that responds preferentially to shape B, as in Fig.3D. The lines and the error bars show the averaged responses and the standard deviations across the population of 10 neurons, respectively.  (C)  The inferred neuronal responses of the 10 sets of V4 neurons, each of which is predicted by a common PFC neuron across degrees of occlusion, projected onto the state space of V4 unit 1 (preferred) and unit 2 (non-preferred) responses. Yellow line represents the averaged inferred responses. Responses to high occlusion are colored green; responses to low occlusion are blue. The left and the right panels show the responses before and after the top-down prediction, respectively. 
}
\label{fig:Population_10}
\end{figure}

The connection weight matrix $\bf{u}$ is learned during the preliminary phase, and the optimal responses of the V4 and PFC neurons with the learned weights $\bf{u}$ are obtained by minimizing the cost function $E$, using the same method as in the previous simulations with single unit representation. Fig.~\ref{fig:Population_10}B shows the averaged inferred responses (dots) and the standard deviation (bars) within the population, of the shape A-selective V4 neurons (green) and the shape B-selective V4 neurons (blue) before (solid line) and after (dotted line) the feedback predictions, as a function of unoccluded area.  Fig.~\ref{fig:Population_10}C illustrates the same results in a state space view, for the responses before (left) and after (right) the feedback. The inferred responses of the neurons in the shape A-selective (V4 unit 1) and the shape B-selective (V4 unit 2) populations, predicted by the common PFC neurons, are projected onto the 2D space of the shape A and shape B-selective population responses. The level of occlusion is indicated by the colorbar, and the yellow line represents the population average responses. The population responses shown in Fig.~\ref{fig:Population_10}B, C match the results from the single-unit representation model in Fig.3 and Fig. 6; the shape discriminability increases during the delayed responses when the feedback predictions are included. Although not shown here, the population average responses of the PFC populations and the occluder-selective V4 population also agree with the previous results in Fig. 3.

Treating each population as a single unit as done in Fig.1 is therefore a reasonable simplification of the model, which expedites computation while maintaining the core mechanisms of the model. Furthermore, there may be recurrent connections among the neurons within the same group, which reduce the variances among these neurons and further validate representation of these neurons as a single unit.

\clearpage

\section*{S2 Appendix. Connection weights learned with partially occluded shapes}

%\beginsupplement

In this appendix, we make predictions on shape discriminability when the synaptic weights store templates of partially occluded shapes instead of unoccluded shapes. This represents the case where the animal has memory of partially occluded shapes rather than being exposed to unoccluded shapes. 

\begin{figure}[hb]%[!htb]
\noindent \centering
\includegraphics[scale=0.7]{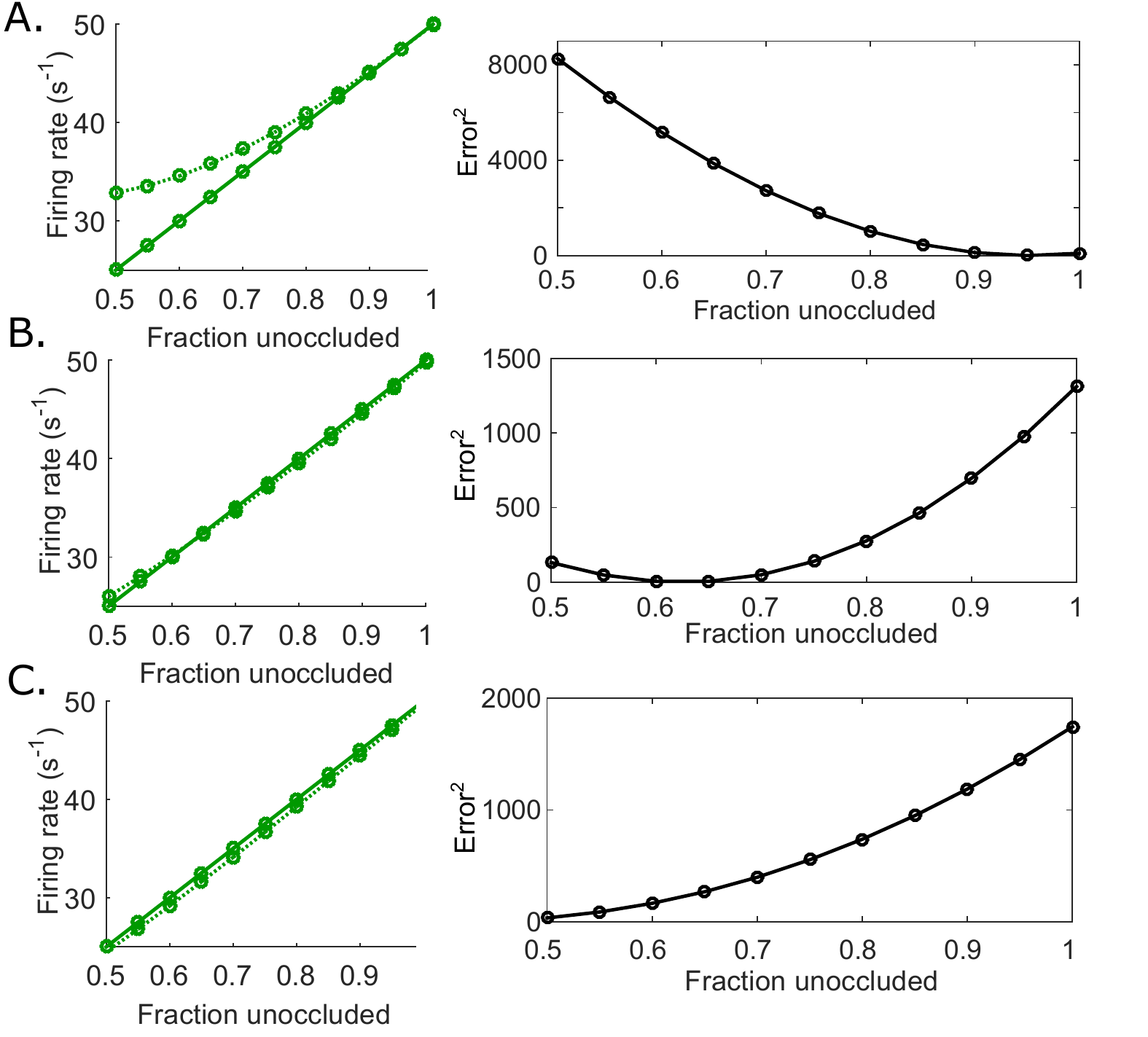}
\caption{{\bf Model simulations when the connection weights are learned from training on partially occluded shapes.}
The initial (solid) and the delayed (dotted) responses of the test shape-selective V4 unit 1 (left column), and the squared total errors between the top-down predictions and the inferred responses of the V4 units (right column), when the connection weight matrix is trained with repeated presentations of (A) unoccluded, (B) $30\%$ occluded, (C) $50\%$ occluded shapes chosen from either shape A or B.}
\label{fig:TrainingWithOcclusion}
\end{figure}

In the simulations in the previous sections, the connection weight matrix $\bf{u}$ is learned based on presentations of the pair of unoccluded shapes, in order to mimic the experimental procedure where the animals discriminated a pair of unoccluded shapes at the beginning of each trial. Here we test our model with the weight matrix learned from partially occluded shapes. We train the weight matrix on the pair of the shapes under $30\%$ and $50\%$  occlusion (Fig.~\ref{fig:TrainingWithOcclusion} B,C) and compare the results to the simulation with the weights learned based on unoccluded shapes (Fig.~\ref{fig:TrainingWithOcclusion} A). 

During the preliminary phase, the gradient descent with respect to the connection weights starts from the initial weight matrix 

\begin{align*}
\bf{u}=\begin{bmatrix}
1& 0.1 \\
0.1 & 1  \\
1 & 1 \\
\end{bmatrix}.
\end{align*}

 Using -1 instead of 0.1 for $u_{2,1}$ and $u_{1,2}$ produces the qualitatively same result. When trained with $30\%$ occlusion, the weight matrix converges to  
 
\begin{align*}
 \bf{u}=\begin{bmatrix}
1.39& 0.48 \\
0.56 & 1.47 \\
2.13 & 2.13 \\
\end{bmatrix},
\end{align*}

and with $50\%$ occlusion, the weight matrix converges to  

 \begin{align*}
 \bf{u}=\begin{bmatrix}
1.27& 0.36 \\
0.29 & 1.19 \\
3.00 & 3.00 \\
\end{bmatrix}. 
\end{align*}

On the other hand, when unoccluded shapes are presented during the training, the weight matrix converges to 

 \begin{align*}
 \bf{u}=\begin{bmatrix}
1.78& 0.85 \\
0.89 & 1.83\\
0.95 & 0.95 \\
\end{bmatrix}.
\end{align*}

When the weight matrix is trained on partially occluded shapes instead of unoccluded shapes, the connection weights to the occluder-selective V4 population converge to larger values over the course of the preliminary training phase. Having the weights learned, the responses of the test-shape preferred V4 unit (V4 unit 1) are plotted across degrees of occlusion, before (solid line) and after (dotted line) the feedback (Fig.~\ref{fig:TrainingWithOcclusion}, left column). We also plotted the total sum of the squared error signals from all three V4 units, namely, the unweighted second term of the cost function  $E$, $\left(\mathbf{r_{v4}}-\bf{u}\cdot\mathbf{r_{pfc}}\right)^T  \left(\mathbf{r_{v4}}-\mathbf{u}\cdot\mathbf{r_{pfc}}\right)$ (Fig.~\ref{fig:TrainingWithOcclusion}, right column). When trained on unoccluded shapes, the squared total error is minimum at zero occlusion. When trained on partially occluded shapes with $30\%$ and $50\%$ occlusion, the squared total error is lowest approximately at the respective occlusion levels (Fig.~\ref{fig:TrainingWithOcclusion}, right column). 

The stronger connection weights between the PFC units and the occluder-selective V4 unit 3, that emerge from training on partially occluded shapes, change the response pattern of the preferred V4. Due to the stronger weights, the PFC responses are relatively lower overall. Then, the delayed responses of shape A-preferred V4 unit 1 induced by PFC predictions are moved to lower values when the stimulus has no or low degrees of occlusion. As the occlusion level increases, the standard deviation $\sigma_1$ of the preferred V4 increases, weakening the bottom-up influence which suppresses the V4 responses under occlusion. As a result, under high occlusion, the optimal representation of the preferred V4 unit 1 responses depends more on the top down PFC prediction reflecting the occluder-selective V4 response pattern.

% The inferred V4 responses with the connection weights trained on occluded shapes show response patterns different from those with the weights trained on unoccluded shapes. 
When the training is based on a pair of unoccluded shapes, the responses of the preferred V4 unit is never lower with the feedback than without the feedback across all occlusion levels, and thus, the feedback enhances the responses under higher degrees of occlusion. On the other hand,  when trained on shapes with $30\%$ occlusion, the delayed responses are lower than the initial responses under low degrees of occlusion. For occlusion levels higher than $\sim 25-30\%$ occlusion, the delayed responses are higher than the initial responses. When trained on $50\%$ occlusion, the delayed V4 responses are always lower than the initial responses in the occlusion range of $0-50\%$. However, the differences between the initial and the delayed responses of the test shape-preferred V4 unit 1 are very small; the initial and the delayed responses are almost identical with the parameter set used here. In addition, across the range of occlusion levels, the errors between the top-down predictions and the inferred V4 activities are smaller when the weights are trained on partially occluded shapes (Fig.~\ref{fig:TrainingWithOcclusion}, right column). 

The deviation from the initial responses (solid line, Fig.~\ref{fig:TrainingWithOcclusion}) is on average smaller for the simulations with weights trained on partially occluded shapes. Since the variance $\sigma_3'$ is smaller than $\sigma_1'$ and $\sigma_2'$,  the prediction $\bf{u}\cdot\mathbf{r_{pfc}}$ tends to follow the response patterns of the occluder-selective V4 unit which increase with added occlusion. When the weights are trained on unoccluded shapes, the connection from a PFC unit to the V4 unit 1 with the same shape preference is the strongest, while its connection to the occluder-selective V4 unit 3 is weaker. Then, the increase in the PFC responses with added occlusion is relatively large, compensating the effects of the small weights $u_{3,1}$ and $u_{3,2}$. The large increase in PFC responses induced by added occlusion can then evoke a larger deviation in the test shape-selective V4 unit 1 from its initial responses. When the weights are trained on partially occluded shapes, compared to the case with training on unoccluded shapes, the weights to the test shape-selective V4 unit 1 are reduced by a little, and the weights to the occluder-selective V4 unit 3 increases significantly. Then, the PFC responses do not increase as much as the occlusion level increase (the preferred PFC unit responses decrease slightly when trained on $50\%$ occlusion or stay constant when trained on $30\%$ occlusion, while the other PFC unit exhibits increasing responses as occlusion increases; data not shown), and thus exert milder effects on the shape-selective V4 units.

In brief, when the connection weights of the network are trained on partially occluded shapes, the feedback from PFC does not improve the shape discriminability. Our model predicts that seeing the unoccluded shapes and learning them prior to the occluded shape-discrimination task may be a necessary step to benefit from the delayed enhancement of shape discriminability induced by the feedback predictions. Testing this hypothesis will be an interesting future experimental study.

\end{document}